%% file: QuAnoTok.tex
\documentclass[11pt, letterpaper]{article}

\usepackage{ifthen}
\usepackage{braket}
\newboolean{delim_grow}
\setboolean{delim_grow}{true}

\input{dgB}

\usepackage{cleveref}
\usepackage{booktabs} 
\usepackage{pifont}
\usepackage{xcolor}
\usepackage{float}   
\usepackage{threeparttable} 
\usepackage{authblk}
\usepackage{tikz}
\usetikzlibrary{shapes,arrows,positioning,fit,calc,decorations.pathmorphing,shadows,decorations.markings}

\usepackage[dvipsnames]{xcolor}
\hypersetup{
	colorlinks=true,
	urlcolor=MidnightBlue,
	linkcolor=MidnightBlue,
	citecolor=MidnightBlue,
	unicode
}

\NewId{\Ps}{\mathcal{P}}
\NewId{\Ms}{\mathcal{M}}

\NewIdR{\Hist}{Hist}
\NewIdR{\Fraud}{Fraud}

\NewIdR{\Mint}{Mint}
\NewIdR{\Report}{Redeem}
\NewIdR{\Test}{Verify}
\NewIdR{\bTest}{\wbr{Verify}}
\NewIdR{\Anon}{Anon}
\NewIdR{\Spy}{Spy}
\NewIdR{\Swap}{Swap}

\NewId{\el}{\eps_\txt{loss}}
\NewId{\dsp}{\delta_\txt{spy}}
\NewId{\gsp}{\gamma_\txt{spy}}
\NewId{\esp}{\eps_\txt{spy}}
\NewId{\ef}{\eps_\txt{fraud}}
\NewId{\NT}{N_{\Test}}
\NewId{\NM}{N_{\Mint}}

\NewOp{\range}{range}

\newcommand{\instDG}{Institute of Mathematics of the Czech Academy of Sciences, \v Zitna 25, Praha 1, Czech Republic}

\newcommand{\DmytroG}{In 2022 the author changed his first name, as explained on his Internet page.}

\newcommand{\thanksDG}{partially funded by the grant 25-16311S of GA \v CR and by RVO:\ 67985840.}

\author[1]{Dmytro Gavinsky\footnote{\DmytroG}}
\author[2]{Dar Gilboa}
\author[2, 3]{Siddhartha Jain}
\author[2]{Dmitri Maslov}
\author[2]{Jarrod R. McClean}

\affil[1]{\instDG}
\affil[2]{Google Quantum AI, Venice, CA, United States}
\affil[3]{The University of Texas at Austin, Austin, TX, United States}

\begin{document}

\title{Anonymous Quantum Tokens with Classical Verification}


\maketitle

\thispagestyle{empty}

\abstart


The no-cloning theorem in quantum mechanics has been used as a basis for quantum money constructions, which guarantee unconditionally unforgeable currency. Existing schemes, however, either (i) require long-term quantum memory and quantum communication between the user and the bank in order to verify the validity of a bill or (ii) fail to protect user privacy due to the uniqueness of each bill issued by the bank, which can allow its usage to be tracked. We introduce a construction of single-use quantum money that gives users the ability to detect whether the issuing authority is tracking them, employing an auditing procedure for which we prove unconditional security. The use of our scheme does not require long-term quantum memory or quantum communication from the users themselves since their validation is a purely classical operation, making the protocol relatively practical to deploy.
We discuss potential applications beyond money, including anonymous one-time pads and voting. 

\abend








\sect[s_intro]{Introduction}

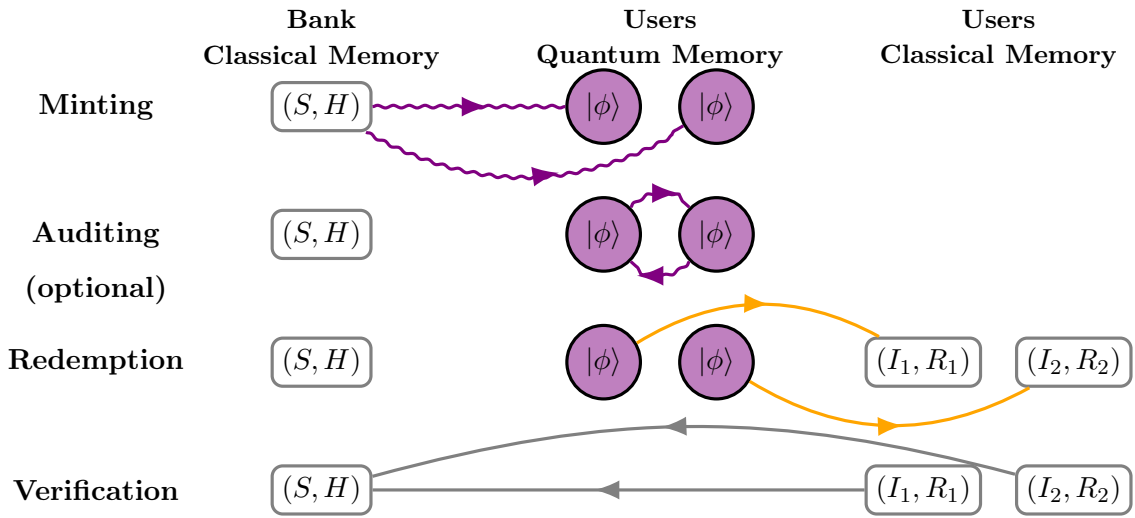
\begin{figure}
    \centering
    \input{tikz_figure}
    \caption{\textit{Anonymous quantum tokens with classical verification.} The four stages of our protocol involve (i) the bank minting quantum tokens and distributing them to users, (ii) users performing optional auditing of the bank to ensure anonymity, (iii) measurement of the quantum tokens, and subsequent storage in classical memory, (vi) validation of the tokens by sending the measurement result to the bank for verification. The pair $(S,H)$ represents a random string used by the bank to generate tokens and the history of redeemed tokens, respectively. Quantum communication/memory is represented in {\bf \textcolor{purple}{purple}}, while classical communication/memory is in {\bf \textcolor{grey}{grey}}. Quantum measurement is represented in {\bf \textcolor{orange}{orange}}.}
    \label{fig:tokens}
\end{figure}

The stochastic and irreversible nature of quantum measurement has no direct classical analog, yet many interesting applications. Among them, it enables one to create objects that are identical before measurement, yet radically distinct after it. Here, we show that this can provide unclonability coupled with indistinguishability, which is useful when one wants to grant controlled access to some resource while guaranteeing the anonymity of the resource users.
We present a construction of tokens that can be used for payment, voting, or other applications, in a way that guarantees their unclonability and the anonymity of the users, while requiring only classical resources for token verification. A schematic representation of our construction is given in Figure \ref{fig:tokens}.

Our work is a novel variant of quantum money~\cite{Wiesner1983-lt, Bennett1983-jx}, an idea dating to the 60's 
which is arguably the first example of exploiting the properties of quantum mechanics towards some useful end (pre-dating the notion of quantum computation), and has spawned a vast literature over the subsequent decades~\cite{Aaronson2009-uz, Gavinsky2011-qd, Aaronson2012-bg, Zhandry2019-ru, Liu2023-pp}. Our scheme involves a bank that distributes suitably structured and identical quantum states to all users. These states can be ``redeemed'' by performing a measurement that produces a classical outcome. Even though initially identical, the result of the measurement of distinct tokens will yield different outcomes with high probability. The resulting classical bit string can be sent to the bank to verify that it indeed originated from a valid token measurement, implying unforgeability. The initial identical nature of the tokens also allows users to compare them using a swap test before measurement, and thus detect whether they are being tracked by the bank.

\newcommand{\cmark}{\textcolor{green!80!black}{\ding{51}}}%
\newcommand{\xmark}{\textcolor{red}{\ding{55}}}%

\begin{table}[ht] 
\small
  \centering
  \begin{threeparttable}
    
    \caption{Comparison of Quantum Money Schemes} \label{tab:quantum_money}
    \renewcommand{\tabcolsep}{3pt}
    \begin{tabular}{l c c c c}
      \toprule
      \textbf{Property} & \textbf{Uncloneability} & \textbf{Cl. verification} & \textbf{Anonymity} & \textbf{Explicit} \\
      \midrule
      Money                                             & \xmark & \cmark & \xmark & \cmark \\
      Quantum Money~\cite{Wiesner1983-lt}               & \cmark & \xmark & \xmark & \cmark \\
      Quantum Coins (Black Box)~\cite{Mosca2010-hc}     & \cmark & \xmark & \xmark\tnote{a} & \xmark \\
      Quantum Coins (Blind QC)~\cite{Mosca2010-hc}      & \cmark & \xmark & \xmark\tnote{a} & \cmark \\
      Quantum Tokens~\cite{Ben-David2023-in}            & \cmark & \cmark & \xmark & \cmark \\
      QM with Classical Verification~\cite{Gavinsky2011-qd} & \cmark & \cmark & \xmark & \cmark \\
      Anonymous Quantum Money~\cite{CGY24}              & \cmark & \cmark & \cmark & \cmark\tnote{b} \\ 
      This Work                                              & \cmark & \cmark & \cmark & \cmark \\
      \bottomrule
    \end{tabular}

    \begin{tablenotes}[flushleft]
      \small 
      \item[a] The constructions of~\cite{Mosca2010-hc} satisfy a certain notion of anonymity only when assuming the honesty of other participants.
      \item[b] Assuming indistinguishability obfuscation.
    \end{tablenotes}

  \end{threeparttable}
\end{table}

One of the main deficiencies of the original quantum money schemes was the need to maintain the quantum money state in memory for long periods of time, which remains technologically prohibitive. While the scheme presented in this work enables the holders of quantum tokens to detect if their tokens are being tracked by the issuer using quantum communication and quantum memory, they may also choose to perform a simple computational basis measurement immediately upon receiving their token. The resulting classical string can then serve for all the applications we discuss, without requiring any additional quantum communication or quantum memory. Nevertheless, the possibility of catching a misbehaving bank can serve as a tangible threat and keep the bank honest, given the presumed cost of the resulting loss of trust. This is akin to other forms of ``expensive scrutiny'' such as a financial audit or health inspection. While the auditing entity may find it expensive to perform audits on all participants, the constant threat of an audit keeps all (or at least most) parties honest. 

Besides maintaining quantum memory for a long period, another drawback of the original scheme by Wiesner~\cite{Wiesner1983-lt} and other subsequent quantum money schemes was the need for quantum communication for verification of a valid money state. This was overcome in~\cite{Gavinsky2011-qd}, and is also implied immediately by the property of our scheme discussed above, since we only require sending the classical string obtained upon measurement and thus do not assume the infrastructure of a quantum internet. Since the implementation of quantum communication faces several technological barriers including signal attenuation~\cite{Van-Meter2014-rt, Leone2022-mq, Devitt2016-oa}, lossy transduction between optical qubits and ones used for computation~\cite{Lauk2020-du, Balram2021-hg, Wang2022-wn}, and the need for fault-tolerant quantum repeaters to communicate across long distances~\cite{Azuma2022-bu}, minimization of quantum communication is a particularly desirable property of any quantum money scheme. Quantum communication between the bank and user is required only in order to issue tokens, in a manner similar to the withdrawal of cash occurring at ATMs, after which the user is free to measure the resulting state immediately, but have it verified by the bank at a later time. 

Recently, \cite{CGY24} introduced an anonymous quantum money scheme that bears some resemblance to our construction, and give several applications including voting. The main advantage of our scheme is that security is unconditional, while \cite{CGY24} require an indistinguishability obfuscation assumption, which is generally considered a strong assumption (as well as another, more well-established hardness assumption). The construction is incomparable to ours since it has advantages with respect it, such as being publicly verifiable (while our tokens can only be verified by the issuer). A comparison of our scheme with previous quantum money constructions is given in \Cref{tab:quantum_money}.

All the reliability claims made in this work (in particular, those regarding the security and the anonymity of the proposed scheme) are \e{resource-independent}:\ i.e., while assuming that the potential adversaries exist in a reality where quantum mechanics is valid, we are not limiting the amount of computational power that they can expend towards attacking our system.
Besides the qualitative advantage in comparison to analogous claims that would depend on the adversary's computational restrictions, our resource-independent guarantees have \e{non-speculative proofs}.\fn
{
  Conditioning the statement on the assumed computational limitations of a potential adversary would likely require relying on unproven hardness assumptions.
}

\ssect[ss_key_results]{Key results}

We define an anonymous quantum token scheme with the following properties:
\begin{itemize}
    \item \textit{Correctness:} Valid tokens pass verification by the bank with high probability, regardless of the behavior of other token holders. 
    \item \textit{Unforgeability:} Given $j$ tokens, the probability that verification is passed more than $j$ times is small.
    \item \textit{Anonymity:} All the tokens are identical in the case of an honest bank, and if the bank tries to modify tokens for traceability, this can be detected by users through the swapping test.
\end{itemize}

Correctness and Unforgeability are defined formally in Definition \ref{d_tok}, and are proved to hold for our scheme in Claim \ref{c_rec_is_tok}. Anonymity is defined and proved in Claim \ref{c_rep_ind}. The formal definition of the construction satisfying these properties is given in Construction \ref{constr_qua}.

Our security proof is based on a generalization of the results that bound the number of input-output pairs of a function that can be reconstructed from quantum queries, first studied by~\cite{FGGS99} and considered in greater generality by~\cite{V13_Cla}. Anonymity of our scheme is essentially a consequence of the tokens being identical, and this property can be checked by users using a swap test without relying on the honesty of the bank.  

\ssect[ss_outline]{Outline of the paper}

The paper is organized as follows.  In Section \ref{s_prel} we present notations, relevant background, and some preliminary results. In Section \ref{s_tokens}, we demonstrate a classical scheme that satisfies unforgeability and correctness but clearly cannot satisfy anonymity, and motivate our quantum construction by first considering some alternative quantum schemes where the anonymity is also violated. Finally, in Section \ref{s_qua_sch}, we present the anonymous token scheme and prove that it satisfies all the desired properties, concluding with the discussion of potential applications in Section \ref{s_app}.


\sect[s_prel]{Preliminaries and notation}
By default, the logarithms are base-$2$.
We will write $[n]$ to denote the set $\set{1\dc n}\sbs\NN$ for $n\in \NN\cup \set{0}$ and let $[a] \deq [\Min{0,\floor a}]$ for $a\in \RR$.
Let $(a,\, b)$, $[a,\, b]$, $[a,\, b)$ and $(a,\, b]$ denote the corresponding open, closed and half-open intervals in $\RR$.
Let $\bt$ and $\tp$ denote, respectively, the false and the true logical (Boolean) values.
We will allow both $\set{\tm}[\tm]$ and $\set[:]{\tm}[\tm]$ to denote sets with conditions (preferring the former).
For a finite set $A$ we will write $\Cl X\unin A$ if the random variable $\Cl X$ is uniformly distributed over $A$.

For a finite $s\sbs\NN$, we will write $s(i)$ to address the $i$'th element of $s$ in natural ordering.
For $x\in\OI^n$, we let $\sz x$ denote its Hamming weight.
For $i\in{[n]}$, we will write either $x_i$ or $x(i)$ to address the $i$'th bit of $x$ (preferring ``$x_i$'' unless it causes ambiguity) and for any $s\sbseq [n]$ both $x_s$ and $x(s)$ will stand for $x_{s(1)}\ds x_{s(|s|)}\in \OI^{|s|}$.

For $f:\: A\to B$ and $A'\sbseq A$, let $f|_{A'}$ denote the restriction of $f$ to $A'$, namely $f|_{A'}:\: A'\to B$ such that $f|_{A'}(x)\equiv f(x)$.
At times we interpret the notation ``$f:\: A\to B$'' as an explicit statement that $f\in B^A$, allowing some natural extensions:\ e.g., ``$f:\: \lrp{A_1\to B_1}\cup\lrp{A_2\to B_2}$'' asserts that $f$ is a function from $A_1\cup A_2$ to $B_1\cup B_2$ such that $f(A_1)\sbseq B_1$ and $f(A_2)\sbseq B_2$.
We will write $g:\: A\leadsto B$ when $g$ may be a non-deterministic operation (usually either classical randomised or quantum by nature) with input in $A$ and output in $B$; the notation described by this paragraph is applicable to such operations too.

For $v\in \CC^n$ or $v\in \RR^n$, let $\norm v$ denote its length.
For a subspace $S$ (respectively denoted $S\le\CC^n$ or $S\le\RR^n$), let $v|_S$ stand for the projection of $v$ onto $S$.

\nfct[f_proj]{Chain inequality for projection}
{
  For $v\in \CC^n$ or $v\in \RR^n$ and subspaces $S_1$ and $S_2$,
  \m{
    \norm{\lf( v|_{S_1} \rt)|_{S_2}}
     \le \norm{v|_{S_2}}.
  }
}

\clm[c_proj_diff_sq]
{
  For $v\in \CC^n$ or $v\in \RR^n$ and subspaces $S_1$ and $S_2$,
  \m{
    \norm{v|_{S_2}}^2 - \norm{\lf( v|_{S_1} \rt)|_{S_2}}^2
     \le 2\tm \norm{v|_{S_1^{\orth}}}\tm \norm{v},
  }
  where $S_1^{\orth}$ is the subspace that complements $S_1$.
}

\prfstart

As
\m{
  v|_{S_2} = \lf( v|_{S_1} + v|_{S_1^{\orth}} \rt)|_{S_2},
}
the triangle inequality implies that
\m{
  \norm{v|_{S_2}} - \norm{\lf( v|_{S_1} \rt)|_{S_2}}
   \le \norm{\lf( v|_{S_1^{\orth}} \rt)|_{S_2}}
   \le \norm{\lf( v|_{S_1^{\orth}} \rt)}
}
and
\m{
  \norm{v|_{S_2}}^2 - \norm{\lf( v|_{S_1} \rt)|_{S_2}}^2
   \le \norm{\lf( v|_{S_1^{\orth}} \rt)}
    \tm \lf( \norm{v|_{S_2}} + \norm{\lf( v|_{S_1} \rt)|_{S_2}} \rt)
   \le \norm{\lf( v|_{S_1^{\orth}} \rt)}
    \tm 2 \tm \norm{v}.
}

\prfend[\clmref{c_proj_diff_sq}]

\ssect[ss_sta_dist]{Quantum states and their distinguishability}[State distinguishability]

We will denote by \Ps[_k] the set of unit vectors in $\CC^{2^k}$, viewed as pure states of $k$ qubits and by \Ms[_k] the set of density matrices in $\CC^{2^k\times2^k}$, viewed as possibly mixed states of $k$ qubits.
For any square $\CC$-matrix $X$, let $\norm{X}[1]$ denote its trace norm $\tr{\sq{X^*X}}$.

\nfct[f_dist]{Distinguishing quantum states}
{
  For any $\sigma_1\ne\sigma_2\in \Ms[_k]$ and a measurement $g:\: \Ms[_k]\leadsto\set{1,2}$:
  \m{
    \PR[i\unin \set{1,2}]{g(\sigma_i) = i}
     \le \fr12 + \fr{\norm{\sigma_1 - \sigma_2}[1]}4
  }
  and the equality is attainable.
  In particular,
  \m{
    \PR[i\unin \set{1,2}]{g(\bk{\phi_i}[]) = i}
     \le \fr12 + \fr{\sq{1-\sz{\bk[\phi_1]{\phi_2}}^2}}2
  }
  for any $\bk{\phi_1}\ne\bk{\phi_2}\in \Ps[_k]$ and the equality is attainable.
}

Very useful for the comparison of a pair of \e{unknown} quantum states can be a measurement called \e{swap test}, due to~\cite{BCWW01_Qua}. We will use the following notation for any $i\in \OI$ and $\sigma\in \Ms[_{2k}]$:
      \m{
        \Swap^{(i)}(\sigma) \deq \PR{\Swap(\sigma) = i}
      }
      and
      \m{
        \wbr{\Swap^{(i)}}(\sigma) \in \Ms[_{2k}]
      }
      will denote the state of the input register after the application of the measurement $\Swap(\sigma)$, subject to having obtained the answer ``$i$''.

\nfct[f_swap]{Swap test~\cite{BCWW01_Qua}}
{
  There exists a measurement $\Swap:\: \Ms[_{2k}]\leadsto\OI$, such that the following holds.
  \enstart
    \item\label{f_swap_pure}
      For any $\bk{\phi},\bk{\psi}\in \Ps[_{k}]$:
      \m{
        \PR{\Swap\lf(\bk{\phi}[\phi]\ox\bk{\psi}[\psi]\rt) = 0} = \fr12 + \fr{\sz{\bk[\phi]{\psi}}^2}{2}.
      }
      In particular, $\Swap(\bk{\phi}[]\ox\bk{\phi}[])\equiv 0$ and the quantum state $\bk{\phi}[]\ox\bk{\phi}[]$ remains unaffected by the measurement.
    \item\label{f_swap_mixed}
      Operationally, the swap test is a two-outcome projection measurement:\ the answer ``$0$'' corresponds to so-called \e{symmetric} subspace and ``$1$'' to its complementary subspace, the \e{anti-symmetric} one.
      That is,
      \m{
        &\bk{\chi}
          = \sq{\Swap^{(0)}(\bk{\chi}[])}\tm \bk{\chi'}{\chi'} + \sq{\Swap^{(1)}(\bk{\chi}[])}\tm \bk{\chi^{\orth}},\\
        &\wbr{\Swap^{(0)}}(\bk{\chi}[])
          = \bk{\chi'}[]\ox\bk{\chi'}[],\\
        &\wbr{\Swap^{(1)}}(\bk{\chi}[])          = \bk{\chi^{\orth}}[]
      }
      for some $\bk{\chi'}\in \Ps[_{k}]$ and $\bk{\chi^{\orth}} \in \Ps[_{2k}]$ such that $(\bk[\chi'][\chi'])\bk{\chi^{\orth}} = 0$.
  \enend
}

In the analysis we will use the following additional properties of the swap test. We first introduce some convenient notation.

\paragraph{Notation for mixed states.} We use ``$\bk{\chi}\in \Ps[_{b+2k}]$'' as a technically convenient reference to an arbitrary mixed state over $2k$ qubits:\ the first $b$ qubits of $\chi$ should be thought of as a ``virtual purification'' of the analysed mixed state (represented by the last $2k$ qubits of $\chi$).

\nclm[c_swap_mixed]{Swap test for mixed states}
{
  For $b, k\in \NN$, let $\bk{\chi}\in \Ps[_{b+2k}]$ and view the $b+2k$ qubits of the state $\bk{\chi}[]$ as three registers:\ call the first $b$ qubits \e{register~0}, the next $k$ qubits \e{register~1} and the last $k$ qubits \e{register~2}.
  Let $\sigma\in \Ms[_{2k}]$, $\sigma_1\in \Ms[_{b+k}]$ and $\sigma_2\in \Ms[_{b+k}]$ be the mixed states corresponding, respectively, to the partial traces of $\bk{\chi}[\chi]$ with traced out register~0, register~2 and register~1 (i.e., $\sigma$, $\sigma_1$ and $\sigma_2$ ``contain'', respectively, the pairs of registers $(1,2)$, $(0,1)$ and $(0,2)$ from $\bk{\chi}[]$).

  Then for any operation $g:\: \Ms[_{b+k}]\leadsto\set{1,2}$:
  \m{
    \PR[i\unin \set{1,2}]{g(\sigma_i) = i}
     \le \fr12 + \sq{\Swap^{(1)}(\sigma)}.
  }
}
Then the statement can be interpreted as follows:
If there is an operation (measurement) that can, given access to register~0, distinguish between registers~1 and~2, then the swap test between registers~1 and~2 can detect the distinction between the registers.

In more technical terms, the claim upper-bounds the probability of distinguishing between $\sigma_1$ and $\sigma_2$ -- both of which are partial traces of $\bk{\chi}[\chi]$ that \e{preserve} the first $b$ qubits -- by a quantity that only depends on $\sigma$, the partial trace of $\bk{\chi}[\chi]$ \e{without} the first $b$ qubits.
What is more, the following proof establishes, in fact, the claimed upper bound with respect to any operation that distinguishes between the state $\bk{\chi}[\chi]$ and the same state with swapped registers~1 and~2.

\prfstart[\clmref{c_swap_mixed}]

Let $\bk{\overleftrightarrow\chi} \in \Ps[_{b+2k}]$ be the state $\bk{\chi}$ where registers~1 and~2 have been swapped.
Obviously, there exists a two-outcome measurement $g'$ that distinguishes between $\bk{\chi}[\chi]$ and $\bk{\overleftrightarrow\chi}[\overleftrightarrow\chi]$ at least as good as $g$ distinguishes between $\sigma_1$ and $\sigma_2$ (e.g., $g'$ may apply $g$ to registers~0 and~1 of the given state) -- accordingly, from \fctref{f_dist}:
\m[m_guess_1]{
  \PR[j\unin \set{1,2}]{g(\sigma_j) = j}
   \le \fr12 + \fr{\sq{1-\sz{\bk[\chi]{\overleftrightarrow\chi}}^2}}2.
}

Consider the Schmidt decomposition of $\bk{\chi}$ with respect to the first $b$ qubits (register~0) vs.\ the rest (registers~1 and~2):
\m{
  \bk{\chi} = \sum_{i=1}^t\lambda_i\bk{\beta_i}{\gamma_i}
}
and let
\m{
  q_i \deq \Swap^{(1)}(\bk{\gamma_i}[]).
}
Then $\sigma = \sum_{i=1}^t\lambda_i^2\bk{\gamma_i}[\gamma_i]$ and
\m[m_sw_dec]{
  \Swap^{(1)}(\sigma)
    = \sum_{i=1}^t\lambda_i^2\tm q_i.
}
According to \fctref{f_swap},
\m{
  \bk{\chi}
    = \sum_{i=1}^t\lambda_i\bk{\beta_i}
      \lf( \sq{1-q_i}\tm \bk{\gamma_i'}{\gamma_i'} + \sq q_i\tm \bk{\gamma_i^{\orth}} \rt).
}

For any $\bk{\gamma} \in \Ps[_{2k}]$ let $\bk{\overleftrightarrow\gamma} \Ps[_{2k}]$ denote the state $\bk{\gamma}$ where the first and the second $k$-qubit registers have been swapped.
Then
\m{
  \bk{\overleftrightarrow\chi}
    = \sum_{i=1}^t\lambda_i\bk{\beta_i}{\overleftrightarrow{\gamma_i}}
    = \sum_{i=1}^t\lambda_i\bk{\beta_i}
      \lf( \sq{1-q_i}\tm \bk{\gamma_i'}{\gamma_i'} + \sq q_i\tm \bk{\overleftrightarrow{\gamma_i^{\orth}}} \rt)
}
and
\m{
  \bk[\chi]{\overleftrightarrow\chi}
    &= \sum_{i=1}^t\lambda_i^2\tm 
      \lf( 1-q_i + q_i\tm \bk[\gamma_i^{\orth}]{\overleftrightarrow{\gamma_i^{\orth}}} \rt)
    \ge 1 - 2\tm \sum_{i=1}^t\lambda_i^2\tm q_i\\
    &= 1 - 2\tm \Swap^{(1)}(\sigma),
}
where the concluding equality is \bref{m_sw_dec}.
Combined with~\bref{m_guess_1}, this gives the required
\m{
  \PR[j\unin \set{1,2}]{g(\sigma_j) = j}
    \le \fr12 + \fr{\sq{1-\sz{\bk[\chi]{\overleftrightarrow\chi}}^2}}2
    \le \fr12 + \fr{\sq{4\tm \Swap^{(1)}(\sigma)}}2.
}

\prfend

\nclm[c_swap_chain]{Chain inequality for swap test}
{
  For $b, k\in \NN$, let $\bk{\chi}\in \Ps[_{b+3k}]$ and view the $b+3k$ qubits of the state $\bk{\chi}[]$ as four registers:\ call the first $b$ qubits \e{register~0} and the following three $k$-qubit tuples \e{register~1}, \e{register~2} and \e{register~3}, respectively.
  For $i<j\in [3]$, let $\Swap_{i,j}(\sigma)$ stand for applying the swap test to registers $i$ and $j$ of $\sigma\in \Ms[_{b+3k}]$ and extend the earlier notation by letting for $\ell\in \OI$
  \m{
    \Swap_{i,j}^{(\ell)}(\sigma) \deq \PR{\Swap_{i,j}(\sigma) = \ell}
  }
  and denoting by $\wbr{\Swap_{i,j}^{(\ell)}}(\sigma)\in \Ms[_{b+3k}]$ the state of the register that initially contained $\sigma$, after the measurement $\Swap_{i,j}(\sigma)$ has been performed and the answer ``$\ell$'' has been obtained.
  Then
  \m{
    \Swap_{1,2}^{(1)}(\bk{\chi}[])
     \le \Swap_{2,3}^{(1)}(\bk{\chi}[]) +
      \Swap_{1,2}^{(1)}\lf( \wbr{\Swap_{2,3}^{(0)}}(\bk{\chi}[]) \rt).
  }
}

Similarly to \clmref{c_swap_mixed}, here we use ``$\bk{\chi}\in \Ps[_{b+3k}]$'' as an effective reference to a mixed state over $3k$ qubits.

Consider a verifier whose goal is to test the claim that \e{the states of registers $1$ and $2$ of $\bk{\chi}[]$ are identical}:\ to \e{accept} with certainty if that is the case and to \e{reject} with the highest possible probability otherwise.
That can be achieved by performing the swap test between the two registers:\ it would detect the difference between the two registers with probability exactly
\m[m_sw_12]{
  \Swap_{1,2}^{(1)}(\bk{\chi}[]).
}

Now imagine that there is an adversary who is willing to ``conceal'' the difference between registers $1$ and $2$.
To do so, he provides another quantum state, as represented by register $3$ of $\bk{\chi}[]$.
In this case the verifier first performs the swap test between registers $2$ and $3$ and rejects right away if the outcome is ``$1$'':\ this happens with probability exactly
\m[m_sw_23]{
  \Swap_{2,3}^{(1)}(\bk{\chi}[]).
}
Otherwise the verifier continues the test as follows:\ he performs the swap test between the registers $1$ and $2$ in their state conditioned upon the outcome ``$0$'' (i.e., non-rejection) in the first swap test; the verifier rejects if the outcome of the second test is ``$1$''; otherwise he accepts.
Then the probability of rejecting after the second swap test equals
\m[m_sw_23_cond]{
  \Swap_{1,2}^{(1)}\lf( \wbr{\Swap_{2,3}^{(0)}}(\bk{\chi}[]) \rt).
}
The the two considered rejection events are mutually exclusive, the total rejection probability equals
\m{
  \Swap_{2,3}^{(1)}(\bk{\chi}[]) + \Swap_{1,2}^{(1)}\lf( \wbr{\Swap_{2,3}^{(0)}}(\bk{\chi}[]) \rt).
}

Intuitively, the two-stage procedure for testing the similarity between registers $1$ and $2$ can be justified as follows:
\itstart
  \item assume that the test has accepted -- i.e., both the swap tests have returned ``$0$'';
  \item then the original states of the registers $2$ and $3$ must have been very similar (as witnessed by the outcome of the first swap test);
  \item accordingly, the first swap test hasn't affected significantly the state of register $2$;
  \item therefore, the second swap test has compared, effectively, the \e{original} state of register $2$ to that of register $1$;
  \item the two states must have been very similar, as witnessed by the outcome of the first swap test, and so the verifier has accepted.
\itend
In accordance to this interpretation, the statement of \clmref{c_swap_chain} can be viewed as a chain inequality for swap test.

\prfstart[\clmref{c_swap_chain}]

The claim follows from a simple fact, as reflected by \fctref{f_swap}:\ the swap test is a projective measurement of its argument.

Consider the two-stage procedure for testing the similarity between registers $1$ and $2$ of $\bk{\chi}[]$, as described above.
Let random variable $\Cl X_1\in \OI$ denote the outcome of the first swap test and $\Cl X_2\in \OI$ the outcome of the second one ($\Cl X_2$ is undefined when $\Cl X_1=1$).
As discussed, the total rejection probability equals
\m[m_swap_chain_0]{
  \PR{\Cl X_1=1 \txt{~or~} \Cl X_2=1}
   = \Swap_{2,3}^{(1)}(\bk{\chi}[]) + \Swap_{1,2}^{(1)}\lf( \wbr{\Swap_{2,3}^{(0)}}(\bk{\chi}[]) \rt).
}
Its complement, the total acceptance probability equals
\m[m_swap_chain_1]{
  \PR{\Cl X_1=0 \txt{~and~} \Cl X_2=0}
   &= \PR{\Cl X_1=0} \tm \PR{\Cl X_2=0}[\Cl X_1=0]\\
   &= \Swap_{2,3}^{(0)}(\bk{\chi}[])
    \tm \Swap_{1,2}^{(0)}\lf( \wbr{\Swap_{2,3}^{(0)}}(\bk{\chi}[]) \rt)
}
-- which is the probability of projecting $\bk{\chi}[]$ first to the subspace corresponding to the outcome ``$0$'' of the measurement $\Swap_{2,3}()$, and then to the subspace corresponding to the outcome ``$0$'' of the measurement $\Swap_{1,2}()$.
By \fctref{f_proj},
\m{
  \Swap_{2,3}^{(0)}(\bk{\chi}[])
      \tm \Swap_{1,2}^{(0)}\lf( \wbr{\Swap_{2,3}^{(0)}}(\bk{\chi}[]) \rt)
    \le \Swap_{1,2}^{(0)}(\bk{\chi}[])
}
and from \bref{m_swap_chain_1}:
\m{
  \PR{\Cl X_1=0 \txt{~and~} \Cl X_2=0}
   \le \Swap_{1,2}^{(0)}(\bk{\chi}[]),
}
that is,
\m{
  \Swap_{2,3}^{(1)}(\bk{\chi}[]) + \Swap_{1,2}^{(1)}\lf( \wbr{\Swap_{2,3}^{(0)}}(\bk{\chi}[]) \rt)
   & = \PR{\Cl X_1=1 \txt{~or~} \Cl X_2=1}\\
   & = 1 - \PR{\Cl X_1=0 \txt{~and~} \Cl X_2=0}\\
   & \ge 1 - \Swap_{1,2}^{(0)}(\bk{\chi}[])\\
   & = \Swap_{1,2}^{(1)}(\bk{\chi}[]),
}
where the first equality is \bref{m_swap_chain_0}.

\prfend

\ssect[ss_]{Limitations of quantum queries}

Let $X$ and $Y$ be finite and $F:\: X\to Y$.
We will call \e{a quantum query} to $F$ an application of a unitary operator $\wht F$ that satisfies the condition
\m{
  \wht F( \bk{x}{\bar 0} ) = \bk{x}{F(x)},
}
where the first register holds an element $x\in X$ and the second register of the result is an element of $Y$
Without loss of generality, the registers consist, respectively, of $\log\ceil{\sz{X}}$ and $\log\ceil{\sz{Y}}$ qubits and some fixed encoding of both $X$ and $Y$ to bit-strings will be implicitly assumed.\fn
{
  We do not specify the output of $\wht F$ when the initial state of the second register differs from $\log\ceil{\sz{Y}}$ qubits in the base-state $\bk0$ in order to avoid discussing the structure of $Y$, which is irrelevant for us now.
}

Let $F:\: X\to Y$ be chosen uniformly at random and an algorithm be allowed at most $q$ queries to $\wht F$.
What would be its success probability of outputting more than $q$ \e{arbitrary} distinct pairs of the form $(x,F(x))$?

For the case when all the generated pairs must be correct for an algorithm to succeed, the question was first investigated by Farhi et al.~\cite{FGGS99} and the answer in the general case is given by~\cite[Theorem 6.1]{V13_Cla}:

\nfct[f_ququ]{\cite{V13_Cla}}
{
  An algorithm making $q$ quantum queries with respect to a uniformly random $F:\: X\to Y$ successfully outputs a list of $r > q$ distinct pairs $(x,F(x))$ with probability at most
  \m{
    \fr 1{\sz{Y}^r} \tm \sum_{i=0}^q \chs ri\tm (\sz{Y}-1)^i.
  }
}


The case when an algorithm is allowed to output a list of $N$ guesses where only $q$ of them are allowed to be correct is analyzed next. This will be useful for the security of our primitive.

\lem[l_ququ]
{
  If an algorithm makes $q$ quantum queries with respect to a uniformly random $F:\: X\to Y$ and outputs at most $N$ distinct pairs $(x_i,y_i)$, then
  \m{
    \PR{\sz{\set{i}[y_i = F(x_i)]} > q} \le 5N\tm \fr{q+1}{\sz{Y}}.
  }
}

Note that $N$ is allowed to be arbitrary, in particular, much larger than $q$; in such a case an algorithm producing an overwhelming majority of ``wrong'' pairs may nevertheless be successful.

\prfstart[\lemref{l_ququ}]

Without loss of generality, assume that $\sz{Y} > 5N$ and $N > q$.

If an algorithm makes $q$ queries with respect to a uniformly random $F:\: X\to Y$ and outputs exactly $q+1$ distinct pairs, then \fctref{f_ququ} implies that the probability that all of them are of the form ``$(x,F(x))$'' -- i.e., the algorithm has guessed all $q+1$ values correctly -- is at most
\m[m_ququ_all_correct]{
  \fr 1{\sz{Y}^{q+1}} \tm \sum_{i=0}^q \chs {q+1}i\tm (\sz{Y}-1)^i
        ~=~ 1 - \fr {(\sz{Y}-1)^{q+1}}{\sz{Y}^{q+1}}
        ~\le~ \fr{q+1}{\sz{Y}}.
}

Let an algorithm $\Cl A_0$ make $q$ quantum queries with respect to a uniformly random $F:\: X\to Y$ and output at most $N$ distinct pairs $(x_i,y_i)$.
Let
\m{
  p_0 \deq \PR{\sz{\set{i}[y_i = F(x_i)]} > q},
}
defined with respect to the output of $\Cl A_0$ with respect to uniformly random $F$.

Let $m_1\deq \floor{\dr{\sz{Y}}{2N}}$ and $F_1$ be a function $X\to [m_1]$.
Consider the following algorithm $\Cl A_1$:
\itstart
  \item Let $\rho_1\dc \rho_{\sz{X}}$ be uniformly random, independently chosen injective functions $[m_1]\to Y$.
  \item Let $\Cl A_0$ run, making $q$ queries to the ``emulated oracle'' $\wht F_1':\: \bk{x}{\bar 0} \to \bk{x}{\rho_x(F_1(x))}$ (the second register consist of $\log\ceil{\sz{Y}}$ qubits).
  \item Let $(x_1,y_1)\dc(x_{N'},y_{N'})$ be the distinct pairs produced by $\Cl A_0$ ($N'\le N$).
  \item Output the content of $\set{\lf(x_i,\rho_{x_i}^{-1}(y_i)\rt)}[y_i\in \range{\rho_{x_i}}]$, truncating it to the first $q+1$ elements if there are more (ordered arbitrarily).
\itend

Let us analyse $\Cl A_1$.
If $\lf(x_i,\rho_{x_i}^{-1}(y_i)\rt)$ produced by $\Cl A_1$ is not of the form $(x,F_1(x))$, then
\m{
  F_1(x_i) \ne \rho_{x_i}^{-1}(y_i)
   ~\Then~
  y_i \ne \rho_{x_i}(F_1(x_i)),
}
but at the same time $y_i\in \range{\rho_{x_i}}$ -- as follows from the fact that the pair has been output by $\Cl A_1$.
This means that the emulated instance of the algorithm $\Cl A_0$ has produced some
\m[m_ququ_y_i]{
  y_i\in \range{\rho_{x_i}},\, y_i \ne \rho_{x_i}(F_1(x_i)).
}

Since for every $x_i$, $\rho_{x_i}$ is chosen uniformly at random and the only element of $\range{\rho_{x_i}}$ seen in the output of the emulated oracle is $\rho_{x_i}(F_1(x_i))$, it is impossible to do better than random guessing to find such a $y_i$ described above.
Accordingly, the probability of finding $y_i$ satisfying \bref{m_ququ_y_i} is at most
\m{
  \fr{\sz{\range{\rho_{x_i}}}-1}{\sz{Y}}
   = \fr{m_1-1}{\sz{Y}}
   \le \fr 1{2N}.
}
As $\Cl A_0$ produces $N'\le N$ distinct pairs, the overall probability of any pair not of the form $(x,F_1(x))$ in the output of $\Cl A_1$ is at most $\dr 12$.

On the other hand, it follows trivially from the construction of $\Cl A_1$ that every element of the form $\lf(x,\rho_x(F_1(x))\rt)$ in the output of $\Cl A_0$ with certainty results in the corresponding $(x,F_1(x))$ appearing in the output of $\Cl A_1$.
Therefore $\Cl A_1$ outputs $q+1$ pairs of the form $(x,F_1(x))$ with probability at least $\dr 12$ times the probability that the output of $\Cl A_0$ making $q$ queries to $\wht F_1'$ contains at least $q+1$ distinct pairs of the form $\lf(x,\rho_x(F_1(x))\rt)$.

Every $F_1:\: X\to [m_1]$ is mapped by the algorithm $\Cl A_1$ to $F_1' = \rho_x(F_1(x))$ which is a uniformly random function $X\to Y$ -- this follows from the uniformly random choice of the injections $\rho_1\dc \rho_{\sz{X}}:\: [m_1]\to Y$.
In other words, the instance of the algorithm $\Cl A_0$ which is used in the construction of $\Cl A_1$ always ``sees'' a quantum oracle $\wht F_1'$ corresponding to a uniformly random $F_1':\: X\to Y$ -- accordingly, its output contains at least $q+1$ distinct pairs of the form $\lf(x,F_1'(x)\rt) = \lf(x,\rho_x(F_1(x))\rt)$ with probability at least $p_0$.

As the earlier analysis of the probability that a pair not of the form $(x,F_1(x))$ appears in the output of $\Cl A_1$ remains valid also when the output of the internal $\Cl A_0$ contains at least $q+1$ distinct pairs of the form $\lf(x,F_1'(x)\rt)$, the algorithm $\Cl A_1$ outputs $q+1$ pairs of the form $(x,F_1(x))$ with probability at least
\m[I]{
  \dr {p_0}2.
}

Finally, it follows from \bref{m_ququ_all_correct} that the probability of $\Cl A_1$ to produce $q+1$ pairs of the form $(x,F_1(x))$ is at most $\fr{q+1}{m_1}$ and therefore,
\m{
  p_0
   \le 2\tm \fr{q+1}{\floor{\dr{\sz{Y}}{2N}}}
   < 5N\tm \fr{q+1}{\sz{Y}},
}
as required.

\prfend

\sect[s_tokens]{Quantum tokens with classical verification\\{\large From traceability to anonymity}}[Tokens with classical verification]

To allow staying in its hotels anonymously, hotel network $\Cl X$ sells ``credit vouchers'':\ every voucher contains a unique secret code that becomes available to the purchaser; a valid secret code can be ``redeemed'' as a payment for a stay in any $\Cl X$-hotel.
As the protocol itself is aimed to provide anonymity, it shouldn't rely on the assumed anonymity of the voucher purchase act (e.g., a personal credit card may be used for that); on the other hand, as the codes are unique, when redeemed, they can be traced back to the person who made the purchase -- thus compromising the desired anonymity.

This metaphoric situation can be addressed by using a quantum cryptographic scheme that we will introduce, formalise and construct in this work.

Somewhat more generally, the main result of this work may be viewed, informally, as an adversary-resistant implementation of the following set-up of \e{anonymous quantum tokens}.
\itstart
  \item The \e{bank} chooses a (private) secret string $\Cl S\unin\OI^{m}$ and uses it to produce $\NM$ identical quantum tokens, which are $k$-qubit pure states.
  We will denote the minting operation by $\Mint(\Cl S)$, it will produce a quantum token (an element of $\Ps[_{k}]$).
  \item An \e{account holder} uses quantum communication (either remotely or via approaching a quantum teller machine) to withdraw some number of quantum tokens -- which can be viewed as \e{single-use quantum banknotes} -- from his bank account into his private quantum wallet.
  \item Tokens from a quantum wallet are used to pay to other account holders -- every transaction requires only classical communication with the bank and the transferred amount gets reflected as the updated balance of the recipient's account at the bank (i.e., every withdrawn token participates in at most one transaction and the corresponding quantum state may be viewed as ``discarded'' afterwards).
  Formally, a measurement denoted by $\Report$ is applied to a token $\bk{\Mint(\Cl S)}$ and the obtained outcome is sent to the bank; the bank either accepts the response and credits the recipient's account or rejects it:\ this is realised by a history-dependent verifying operation $\Test()$ (cf.\ \defiref{d_tok} and the following discussion).
  \item The scheme guarantees safety (i.e., future usability) of yet unused withdrawn tokens, even when other account holders adversarially cooperate against it -- except with negligible probability.
  \item The scheme guarantees tokens' counterfeit resistance (even against cooperated adversarial efforts of all account holders), except with negligible probability.
  \item The scheme doesn't allow payer's identity tracing by an adversarial bank, except with negligible probability.
\itend

Note, in particular, that the ``anonymous hotel stay'' scenario can be implemented via an anonymous quantum tokens scheme where the hotel network plays the roles of both the bank and the transaction recipient.

To formalise this concept, we shall first consider a non-anonymous version (informal discussion will follow the definition).

\ndefi[d_tok]{Quantum tokens with classical verification}
{
  We call a tuple
  \m{
    \lrp{\Mint(), \Report(), \Test()}
  }
  an \e{$(n, m, \NM, \NT, t, \el, \ef)$-scheme with classical verification} if the following holds:\fnm
  \enstart
      \item\label{d_tok_test_pass}
      $\Mint:\: \OI^{m}\leadsto (\Ps[_{n}])^{\NM}$, $\Report:\: \Ms[_{n}]\leadsto\OI^{t}$ and $\Test:\: \OI^{m}\times \bigcup_{j=0}^{\NT-1}(\OI^t)^j\times \OI^{t} \leadsto \TF$.
    \item\textit{(Correctness)}
      Let ``$\bk{\Mint[_i](\Cl S)}$'' denote the $i$'th $n$-qubit register of $\Mint(\Cl S)$, then for any $0\le j_0\le\NT-1$, $i_0\in [\NM]$ and a measurement \Hist\ of a quantum state over $n\tm(\NM-1)$ qubits with outcomes in $\OI^{t\tm j_0}$:
      \m{
        \PR[\Cl S, \Cl H, \Cl R]{\Test(\Cl S, \Cl H, \Cl R) = \tp} \ge 1-\el,
      }
      where $\Cl S\unin\OI^{m}$ is the bank's secret string, $\Cl H$ is the outcome of the measurement \Hist\ applied to the state $\bk{\Mint[_1](\Cl S)}\ds\bk{\Mint[_{i_0-1}](\Cl S)}{\Mint[_{i_0+1}](\Cl S)}\ds\bk{\Mint[_{\NM}](\Cl S)}$ and $\Cl R$ is the outcome of $\Report(\bk{\Mint[_{i_0}](\Cl S)}[])$.

    \item\label{d_tok_fraud}\textit{(Security/Unforgeability)}
      For any $0\le j\le\NT$ and $h\in (\OI^t)^j$, let $\bTest(s, h)$ denote the bit-string $a\in \OI^j$ where $\forall\: i\in [j]$:\ $a_i \teq \Cases{1}{if $\Test(s, h_{\{0\dc (i-1)t\}}, h_{\{(i-1)t+1\dc it\}}) = \tp$;}{0}{otherwise.}$
      Then for any $s_0\sbseq [\NM]$ and a measurement \Fraud\ of a quantum state over $n\tm\sz{s_0}$ qubits with outcomes in $\OI^{\NT\tm t}$:
      \m{
        \PR[\Cl S, \Cl H]{\sz{\bTest(\Cl S, \Cl H)}>\sz{s_0}} \le \ef,
      }
      where $\Cl S\unin\OI^{m}$ is the bank's secret string and $\Cl H$ is the outcome of the measurement \Fraud\ applied to the state $\bk{\Mint[_{s_0(1)}](\Cl S)}\ds\bk{\Mint[_{s_0(\sz{s_0})}](\Cl S)}$.
  \enend
}
\fnt[fn_par_n]{Every element specifying a scheme may be parametrised by the standard \e{complexity parameter} $n$ or its analogue.
The probabilities in the definition are taken with respect to both the listed random variables and the operational randomness of the considered procedure.}

Intuitively, in \defiref{d_tok} the parameter $n$ denotes the number of qubits per token; $m$ is the bit-length of bank's secret string (denoted by $\Cl S$ -- chosen uniformly at random from $\OI^{m}$); $\NM$ is the number of tokens that are issued (using the value of the secret string $\Cl S$ that the bank has picked); $\NT \ge \NM$ is an upper bound on the total number of times that the bank's validity test (corresponding to the same secret string) is allowed to be initiated; $t$ is the number of bits in every message sent to the bank towards (classical) token verification; $\el$ bounds the probability that an authentic properly stored token doesn't pass bank's verification procedure (in the worst-case scenario when the holders of all other tokens act adversarially); $\ef$ bounds the probability that holders of $s\le \NM$ tokens who act adversarially are able to pass bank's verification procedure at least $s+1$ times by making at most $\NT$ attempts.
The operation $\Mint()$ is the token-minting procedure; $\Report()$ is the measurement operation applied by a holder to his token in order to generate the verification message for the bank; $\Test()$ is the bank's verification procedure.

Note that bank's verifying action is \e{non-interactive} (that will suffice for the goals of this work) but, in general, \e{history-dependent}:\ namely, the input to $\Test()$ consists -- besides the bank's private string $\Cl S$ and the $t$-bit tuple $\Cl R$ that has been submitted for the current verification procedure -- of up to $\NT-1$ $t$-bit tuples representing the bit-strings that were submitted for verification earlier, in the sequence of their arrival (denoted by $\Cl H$).
From the operation standpoint, one may think about the value of $\Cl H$ as a dynamic database maintained by the bank, where all submissions to the verifying protocol are being stored, while the value of $\Cl S$ is stored as a static value. 
The value returned by $\Test(\Cl S, \Cl H, \Cl R)$ is \e{accept} ($\tp$) if the verification has been passed and \e{reject} ($\bt$) if failed.

Let us call a \e{minted series} the collection of $\NM$ tokens corresponding to the same bank's secret string $\Cl S$ -- i.e., the collection of $\NM$ $n$-qubit registers in $\Mint(\Cl S)$.
Then condition~\ref{d_tok_test_pass} of \defiref{d_tok} can be interpreted as a requirement that the unused token corresponding to $\bk{\Mint[_{i_0}](\Cl S)}$ remains usable -- i.e., gets accepted by the bank's verification procedure $\Test()$ -- with probability at least $1-\el$, even if the holders of the other tokens of the same minted series $\Mint(\Cl S)$ ``cooperate against'' it:\ i.e., they apply the measurement $\Hist$ to all other tokens in order to generate an adversarial verification history $\Cl H$ that is aimed to trick the bank into rejecting $\bk{\Mint[_{i_0}](\Cl S)}$.\,\fn
{
  One cannot allow the bank to be adversarial in the context of token safety:\ otherwise the model would allow no non-trivial implementation, as an adversarial bank can simply reject all the minted tokens.
}

Condition~\ref{d_tok_fraud}, on the other hand, can be interpreted as a guarantee that the scheme is counterfeit-resistant -- i.e., no effort of the holder(s) of the tokens indexed by $s_0\sbseq [\NM]$ from the minted series $\Mint(\Cl S)$ would allow them to pass more than $\sz{s_0}$ verifications by bank with probability more than $\ef$, as long as the total number of attempts that they can make is at most $\NT$.
Here the possible adversarial efforts are represented by the generic procedure \Fraud: it measures the $\sz{s_0}$ tokens (note that $[\sz{s_0}=0]$ is allowed) and generates the input for approaching the bank's verification procedure $\NT$ times -- $\bTest(\Cl S, \Cl H)$ is the corresponding sequence of $\NT$ bank's responses (to make the adversary ultimately powerful, we let all the verification requests that the bank receives come from the output of \Fraud\ and the total number of the requests be the maximum allowed, namely $\NT$).

Viewing a hypothetical classical construction as a special case of the above definition, we can ask right away whether a \e{classical} token scheme with classical verification exists.
It is not hard to see that the hotel-credit vouchers being issued by the network $\Cl X$ (as described in the beginning of this section) do implement such a scheme.

\nconstr[constr_cla]{a classical token scheme}
{
  Let $4|k$, $\NM\deq 2^{\dr k4}$, $m\deq k\tm\NM$, $\NT\deq 2^{\dr k2}$ and $t\deq k$.
  For simplicity and convenience, let us use $\OI^k$ instead of $\Ps[_k]$ in the context of a classical construction.
  The operations $\Mint:\: \OI^{m}\to (\OI^k)^{\NM}$, $\Report:\: \OI^k\to\OI^{t}$ and $\Test:\: \OI^{m}\times \bigcup_{j=0}^{\NT-1}(\OI^t)^j\times \OI^{t} \to \TF$ are defined as
  \m{
    \Mint(S)&\deq \lrp{S(\set{1\dc k}),\, S(\set{k+1\dc 2k})\dc\, S(\set{m-k\dc m})};\\
    \Report(\tau)&\deq \tau;\\
    \Test(S, H, R)&\deq \Cases{\tp}{if $R \in \Mint(S)$ and $R \nin H$;}{\bt}{otherwise.}
  }
}

To analyse \constrref{constr_cla}, let $S_0$ be the value taken by the random variable $\Cl S\unin\OI^{m}$.
Denote by $S_0^{(i)}$ for $i\in [\NM]$ the $k$-bit string $S_0(\set{(i_0-1)k+1\dc i_0k})$ -- that is, the $i$'th register of $\Mint(S_0)$.

First we verify that condition~\ref{d_tok_test_pass} of \defiref{d_tok} holds and compute the corresponding value of $\el$.
That is, for some $i_0\in [\NM]$ we want to lower-bound the probability of the event $[\Test(\Cl S, \Cl H, \Cl R) = \tp]$, where, according to the above definitions, $\Cl R \equiv S_0^{(i_0)}$ and $\Cl H$ consists of $j_0\le\NT-1$ $k$-bit registers, which are all independent from $S_0^{(i_0)}$ (as it may only depend on $S_0^{(1)}\dc S_0^{(i_0-1)}, S_0^{(i_0+1)}\dc S_0^{(\NM)}$ and $\Cl S\unin\OI^{m}$).
As $S_0^{(i_0)}\in \Mint(S)$ by construction, the event $[\Test(\Cl S, \Cl H, \Cl R) = \bt]$ can happen only if $R=S_0^{(i_0)} \in \Cl H$ -- that is, if at least one of the $k$-bit registers of $\Cl H$ matches $S_0^{(i_0)}$.
By the union bound, that happens with probability less than $\el\deq \NT\tm2^{-k} = 2^{-\dr k2}$.

Note that \constrref{constr_cla} allows the situation when $\Mint(S)$ produces several \e{identical} token values:\ in this case all but one (namely, the first one submitted for verification) repeated tokens will be rejected by the bank.
We intentionally leave the probability of this event positive (as well as the probability of an analogous event in our main construction later on):\ on the one hand, it is less than $\el=2^{-\dr k2}$, as follows from the above analysis; on the other hand, disallowing such events would break the convenient symmetry resulting from the perfect mutual independence of the tokens, thus making the analysis of the scheme more cumbersome.

Now consider condition~\ref{d_tok_fraud} of \defiref{d_tok}.
Here for any $s_0\sbseq [\NM]$ we want to upper-bound the probability that $\sz{\bTest(\Cl S, \Cl H)}>\sz{s_0}$, where $\Cl H$ is a random variable consisting of $\NT$ $k$-bit registers that may depend only on $(S_0^{(i)})_{i\in s_0}$ among the $\NM$ registers of $\Mint(S_0)$.
By the above construction, $\sz{\bTest(\Cl S, \Cl H)}>\sz{s_0}$ only if the content of at least one of the registers $S_0^{(i)}$ for $i\nin s_0$ appears among the $\NT$ registers of $\Cl H$ -- by the union bound, that happens with probability at most $\NT\tm(\NM-\sz{s_0})\tm2^{-k} < 2^{-\dr k4}$; accordingly, the condition holds with respect to $\ef\deq2^{-\dr k4}$.

Accordingly, \constrref{constr_cla} is a valid token scheme with classical verification that satisfies \defiref{d_tok} with parameters $\combr \big(n=k, m=k\tm2^{\dr k4}, \NM=2^{\dr k4}, \NT=2^{\dr k2}, t=k, \el=2^{-\dr k2}, \ef=2^{-\dr k4}\big)$ for any $4|k$.
These parameters are close to the best possible.

\ssect[ss_tow_anon]{Towards anonymity: Repetitive tokens}[Repetitive tokens]

Any non-degenerate classical implementation of \defiref{d_tok} is necessarily non-anonymous in the following sense:\ to be counterfeit-resistant, the tokens produced by $\Mint()$ must be unique (except with negligible probability) and such tokens can be traced by the bank when they are being verified.
The same high-level argument doesn't apply when the token scheme is quantum:\ in particular, simultaneous circulation of multiple identical quantum tokens does not necessarily imply susceptibility to counterfeit, as quantum states cannot, in general, be duplicated (due to the quantum uncertainty).

Before formalising our notion of anonymity, let us have a look at some examples of non-anonymous -- that is, traceable by the bank -- quantum token constructions.

Consider the following simple set-up:\ ideally, the bank uses its secret private string $S_0\in \OI^m$ to generate multiple identical copies of the token state
\m[m_fair]{
  \sum_{i\in [m]}\bk{i}{S_0(i)}
,}
towards verification a token holder measures this state in the computational basis and sends the outcome $(i, a)$ to the bank; the verification procedure $\Test(S_0, H, (i, a))$ accepts if and only if both $S_0(i) = a$ and the pair $(i, a)$ has not yet been submitted for verification -- i.e., $(i, a) \nin H$.\,\fn
{
  This ``toy example'' is, obviously, rather primitive:\ in particular, a cheater can easily pass such verification procedure with probability almost $\dr12$; nevertheless, our main \constrref{constr_qua} will share some similarities with this one.
}

Now suppose that the bank is willing to trace a specific user and instead of a fair token it supplies him with the register~$2$ of the state
\m[m_loaded]{
  \sum_{i\in [m]}\overbrace{\bk{i}}^{1}\overbrace{\bk{i}{S_0(i)}}^{2}
,}
while register~$1$ remains in the bank.
Then the distribution of the traced user's messages submitted for verification is exactly the same as that distribution for a holder of a ``fair'' (presumably untraceable) token~\bref{m_fair}; nevertheless, the bank is able to recognise, with high probability, a call for verification originating from the ``loaded'' (traced) token, as the value of $i$ in that message would equal the value of $i$ in the first register of the state~\bref{m_loaded}.

Here is another example.
Under the same token scheme, let the bank choose uniformly at random a permutation $h:\: [m]\to[m]$ and provide the targeted client with two tokens in the form of
\m[m_loaded2]{
  \sum_{i\in [m]}\overbrace{\bk{i}{S_0(i)}}^{1}\overbrace{\bk{h(i)}{S_0(h(i))}}^{2}
,}
where the registers~1 and~2 represent, respectively, the first and the second client's token.
As long as the number of past verifications is within \aso{\sq m} (e.g., when $\NT=\sq[3] m$), the appearance in the verification history of both $(i_0, S_0(i_0))$ and $(h(i_0), S_0(h(i_0)))$ would likely mean that those two requests have arrived from the client being spied on.
Unlike~\bref{m_loaded}, example~\bref{m_loaded2} demonstrates a concealed surveillance set-up which doesn't rely upon any entanglement of the ``loaded'' tokens with the bank.

These examples emphasise that the sole requirement that the distribution of the messages submitted for bank's verification is the same for all circulating tokens is insufficient for guaranteeing the desired anonymity of the scheme.

Intuitively, a token scheme would be perfectly anonymous if the bank were ``supernaturally'' forced to produce a minted series consisting of $\NM$ \e{identical} pure states.
Let us define schemes were the \e{legitimate} minting action is like that.

\ndefi[d_tok_rep]{Repetitive quantum tokens with classical verification}
{
  We call a tuple
  \m{
    \lrp{\Mint(), \Report(), \Test()}
  }
  a \e{repetitive $(n, m, \NM, \NT, t, \el, \ef)$-scheme with classical verification} if it is a $\combr(n, m, \NM, \NT, t, \el, \ef)$-scheme (cf.~\defiref{d_tok}) and in addition the following holds:
  \m{
    \forall\: s\in \OI^{m}:\: 
      \bk{\Mint(s)} \equiv \bk{\Mint'(s)}^{\ox\NM}
  }
  for some $\Mint':\: \OI^{m}\leadsto\Ps[_{n}]$.
}

There is a manifest difference between \defiref[d_tok]{d_tok_rep}:\ the former can be viewed as ``rather semantic'', directly addressing the relevant security aspects of the defined objects and operations; the latter, on the other hand, is ``purely syntactic'' (besides its reference to \defiref{d_tok}), ignoring any operational impact of the additional constraint on $\Mint()$.
While the constraint is aimed to imply anonymity, it is not yet clear how to impose it on the potentially adversarial bank aiming to track its clients' activities.

In other words, if we address (somewhat informally) the guarantees of \defiref{d_tok} as \e{the security} of a token scheme -- as opposed the \e{the anonymity} of it, which is handled by \defiref{d_tok_rep} -- then we can say that the security guarantees of our cryptographic primitive are semantically defined, while the anonymity guarantees would follow from certain syntactic properties of a construction satisfying the definition.

The ``non-operational'' flavour of \defiref{d_tok_rep} offers some benefits.
\itstart
  \item The security guarantees are essentially the same for every possible application (a cryptographic protocol) that uses our primitive:\ namely, it is (unconditionally) infeasible to try and generate $k+1$ valid verification messages by measuring at most $k$ tokens for any $k \le \NM$.
  On the other hand, the precise notion of user anonymity may depend on the particular application (e.g., an adversarial bank may be interested either in the exact identification of the account holder behind a specific ``targeted operation'' or in collecting more general ``statistics'' of user activities; moreover, we will see in \Cref{s_app} that our \defiref{d_tok_rep} can be used for constructing non-token schemes, like voting or certain forms of one-time pads -- those applications have their own notion of anonymity).
  An attempt to formalise semantically the generic anonymity guarantees that would cover all possible cases of interest would likely lead to an overwhelmingly complex definition; instead, we give a very generic syntactic definition, thus ``delaying the case analysis'' to the actual anonymity proofs (of which we will present some typical examples, easily adaptable to the specific anonymity requirements).
  \item Our simple syntactic definition allows more flexible and ad hoc anonymity proofs 
  -- thus shorter and simpler technically.
  This simplicity, in turn, will lead to more direct and less prone to logical errors formulations of the cryptographic guarantees.
  \item Besides the technical merits, the concise formulation of \defiref{d_tok_rep} makes it, in our opinion, more intuitive.
  Unlike the general tokens (\defiref{d_tok}), the repetitive ones cannot be implemented classically in any meaningful way and \defiref{d_tok_rep} concisely highlights the quantum essence of the possible anonymity of a token scheme.
\itend

If a token scheme satisfies \defiref{d_tok_rep}, then any detected discrepancy between a pair of tokens from the same minted series can be interpreted as a protocol violation by the token-issuing authority (potentially target against user anonymity).
Intuitively, in this case one of the most useful approaches that can be taken by the users to test bank's trustworthiness would be to perform the swap tests between some randomly picked tokens, originally held by random users:\ if the bank has deviated from the protocol in order to be able to distinguish between a pair of tokens (say, coming from different users), then the swap test over the same pair would detect that the tokens are not identical; moreover, we can expect the probability of the detection to be roughly proportional to the success probability of the bank's user-tracking activities.

\ssect[ss_anon]{Anonymity of repetitive schemes}

Now we give a general framework for deriving anonymity guarantees of cryptographic applications based upon our repetitive tokens.

Recall that a scheme implementing \defiref{d_tok} can be thought as the underlying cryptographic primitive for the following scenario:
\itstart
  \item The bank draws at random a secret bit-string $\Cl S$, which is then used with the operation $\Mint()$ to produce (mint) the tokens.
  \item An account holder withdraws from the account quantum tokens $\sigma_1\dc \sigma_\ell$.
  \item To make payments, the holder uses his tokens one by one as follows:\ first the measurement $\Report()$ is applied to a token; then the obtained answer is sent to the bank for verification and the token (quantum state) that has been used gets discarded.
  \item To verify a message received for token verification, the bank uses the operation $\Test()$ in combination with its secret bit-string $\Cl S$ and answers accordingly.
\itend
In addition to this, a repetitive scheme (\defiref{d_tok_rep}) can be used to ensure users anonymity:\ e.g., as follows.
\itstart
  \item Before starting spending his tokens, the account holder obtains a random \e{pattern token} $\sigma_\txt{pat}$ from some other user -- usually this is done only once.\fn
{
  Although the details of a protocol for obtaining a random ``outside'' token by an account holder is beyond the scope of our model, let us note that it is conceivable that the users may be willing to perform randomised ``token exchange'' once in a while, as verifying bank's non-intrusion is in their mutual interest.
}
  \item To make payments, the user first applies the swap test to the pair of tokens $\sigma_{i_0}$ and $\sigma_\txt{pat}$, and only if it returns ``$0$'' (cf.~\fctref{f_swap}), our user proceeds by applying the measurement $\Report()$ to $\sigma_{i_0}$ and sending the obtained answer to the bank -- in this case the pattern token $\sigma_\txt{pat}$ is preserved for future verifications.
  If, on the other hand, the swap test returns ``$1$'' (thus witnessing inequality between the tokens), then the user knows that he is being watched by the bank and acts accordingly.
\itend

In this part we shall formalise the anonymity argument.
The analysis will be based on \defiref{d_tok_rep} alone -- i.e., ignoring any particular properties of concrete implementations of a repetitive token scheme.

When the operation $\Report:\Ms[_n]\leadsto\OI^{t}$ is clear from the context, let us denote by $\Report':\Ms[_{2n}]\leadsto\OI^{t}\cup \set{\bot}$ the following operation ($\sigma\in \Ms[_{2n}]$ is the input state):
\itstart
  \item Perform $\Swap(\sigma)$, abort the operation and output ``$\bot$'' if the swap test returns ``$1$'' (report that cheating has been detected).
  \item For $\wtl{\sigma}|_{n+1\dc2n}$ denoting the last $n$ qubits of the input state after performing the swap test, perform $\Report(\wtl{\sigma}|_{n+1\dc2n})$ and output the response.
\itend

To begin, let us see that \defiref{d_tok_rep} corresponds to a token scheme in the following sense.

\clm[c_rec_is_tok]
{
  Let $\lrp{\Mint(), \Report(), \Test()}$ for $\bk{\Mint(s)} \equiv \bk{\Mint'(s)}^{\ox\NM}$ be a repetitive $(n, m, \NM, \NT, t, \el, \ef)$-scheme with classical verification (cf.~\defiref{d_tok_rep}).
  Then for $\wbr\Report(\tau)$ defined as $\Report'(\bk{\Mint'(s_0)}[]\ox \tau)$ where $s_0$ is the bank's secret string,
  \m{
    \lrp{\Mint(), \wbr\Report(), \Test()}
  }
  is a $\combr(n, m, \NM, \NT, t, \el, \ef)$-scheme (cf.~\defiref{d_tok}).
  If $\Report()$ has an efficient implementation, then $\wbr\Report()$ has one as well.
}

\prfstart

In the beginning of the operation every user selects a \e{pattern token} $\tau_\txt{pat}$ among his tokens (towards anonymity the users would like to exchange their patter tokens randomly, which we will discuss later).
The swap test in the beginning of $\wbr\Report()$ corresponds to $\Swap(\tau_\txt{pat}\ox \tau)$ -- i.e., comparing the pattern token to the token used in the transaction.

As the swap test can be performed efficiently, $\wbr\Report()$ has an efficient implementation if $\Report()$ has it.\fn
{
  When only the pattern token remains at the user's disposal, he skips the swap test in the beginning and outputs $\Report(\tau_\txt{pat})$ right away.
}

As long as the bank is acting according to the requirements of \defiref{d_tok_rep}, all tokens in the minted series are identical, and therefore $\Swap(\tau_\txt{pat}\ox \tau)$ returns ``$0$'' with certainty and doesn't affect the state of its input -- both $\tau_\txt{pat}$ and $\tau$ stay intact.
In particular, $\Report(\wtl{\tau})$ produces the same output distribution as $\Report(\tau)$ and the result follows.

\prfend[\clmref{c_rec_is_tok}]

The swap test in the first step of $\Report'()$ is meant to ensure the anonymity of a repetitive scheme.


\clm[c_rep_ind]
{
  For $b\in \NN$ and $\lrp{\Mint(), \Report(), \Test()}$ a repetitive $(n, m, \NM, \NT, t, \el, \ef)$-scheme with classical verification for $\bk{\Mint(s)} \equiv \bk{\Mint'(s)}^{\ox\NM}$, let $\bk{\chi}\in \Ps[_{b+2n}]$ and call the first $b$ qubits of $\bk{\chi}[]$ \e{register~0}, the next $n$ qubits \e{register~1} and the last $n$ qubits \e{register~2}.

  For $i\ne j\in \set{0,1,2}$, denote by $\sigma_{i}$ the partial trace of $\bk{\chi}[]$ that ``keeps'' the qubits of register~$i$ and by $\sigma_{i,j}$ the partial trace (possibly followed by qubit-permutation) that corresponds to the qubits of register~$i$ followed by those of register~$j$.
  For $i\in [2]$, denote by $\Cl X_i$ the output of $\Report(\sigma_{i})$ and by $\Cl X_{2,1}$ that of $\Report'(\sigma_{2,1})$.

  Then for any operation $g:\: \Ms[_{b}]\times\OI^{t} \leadsto\set{1,2}$ and $a_1 = 1$ and $a_2\in \set{2,(2,1)}$ the following holds:
  \m{
    \PR[k\unin \set{1,2}]{g(\sigma_0,\Cl X_{a_k}) = k}
     \le \fr12 + \sq{ \PR{\Cl X_{2,1} = \bot} }.
  }
}

Here we use ``$\bk{\chi}\in \Ps[_{b+2n}]$'' as a technical reference to an arbitrary mixed state of a pair of user-held tokens -- represented by registers~1 and~2, respectively -- with the maximum possible control by the bank, which holds register~0.

Intuitively, the claim asserts that if the bank violates the scheme in a way that allows it either distinguishing between the usage of two tokens in a transaction or detecting whether the swap test between a pair of tokens was used before applying the operation $\Report()$ to one of them 
, then the same swap test will likely detect this breach of anonymity.

\prfstart[\clmref{c_rep_ind}]


For $a_1=1$ and $a_2=2$, let $g':\: \Ms[_{b+n}]\leadsto\set{1,2}$ be the following operation ($\sigma$ is the input state):
\itstart
  \item Perform $\Report()$ over the last $n$ qubits of $\sigma$, denote the response by $r$.
  \item Perform $g(\sigma|_{[b]}, r)$, where $\sigma|_{[b]}$ denotes the first $b$ qubits of $\sigma$, and output the response.
\itend
Then
\m{
  \PR[k\unin \set{1,2}]{g(\sigma_0,\Cl X_{a_k}) = k}
   & = \PR[k\unin \set{1,2}]{g'(\sigma_{0,k}) = k}\\
   & \le \fr12 + \sq{\Swap^{(1)}(\sigma_{1,2})}
    = \fr12 + \sq{ \PR{\Cl X_{2,1} = \bot} },
}
where the inequality is \clmref{c_swap_mixed}.

For $a_1=1$ and $a_2=(2,1)$, let $h:\: \Ps[_{b+2n}]\leadsto\set{1,2}$ be the following operation ($\bk{\chi}$ is the input state):
\itstart
  \item Perform $\Report()$ over the last $n$ qubits of $\bk{\chi}[]$, denote the response by $r$.
  \item Perform $g(\bk{\chi}[]|_{[b]}, r)$, where $\bk{\chi}[]|_{[b]}$ denotes the first $b$ qubits of $\bk{\chi}[]$, and output the response.
\itend
Without loss of generality, there is a two-outcome projective measurement represented by $S_h^{(1)}\orth S_h^{(2)}\le \CC^{2^{b+2n}}$, such that the outcome $i\in \set{1,2}$ of $h(\bk{\chi})$ corresponds to projecting the vector $\bk{\chi}$ to the subspace $S_h^{(i)}$.

The output of $h(\bk{\chi})$ -- that is, of the measurement $S_h^{(1)}$ vs.\ $S_h^{(2)}$ -- is distributed like that of $g(\sigma_0,\Cl X_1)$.
On the other hand, the output of $g(\sigma_0,\Cl X_{2,1})$ is distributed like that of the swap test performed on the last $2n$ qubits of $\bk{\chi}[]$, in the case of the outcome ``$0$'' followed by performing the measurement $S_h^{(1)}$ vs.\ $S_h^{(2)}$.
As the swap test is a projective measurement to the symmetric vs.\ the anti-symmetric subspace (cf.\ \fctref{f_swap}), from \clmref{c_proj_diff_sq} it follows that
\m{
  \PR{g(\sigma_0,\Cl X_1) = 1} - \PR{g(\sigma_0,\Cl X_{2,1}) = 1}
   &\le 2\tm \sq{ \Swap^{(1)}(\bk{\chi}[]|_{b+1\dc b+2n}) }\\
   &= 2\tm \sq{ \PR{\Cl X_{2,1} = \bot} },
}
where $\bk{\chi}[]|_{b+1\dc b+2n}$ are the last $2n$ qubits of $\bk{\chi}[]$.
Accordingly,
\m{
  \PR[k\unin \set{1,2}]{g(\sigma_0,\Cl X_{a_k}) = k}
   \le \fr12 + \sq{ \PR{\Cl X_{2,1} = \bot} }.
}

\prfend

To conclude our analysis of the core anonymity properties of a repetitive quantum token scheme with classical verification, let us see that the same pattern token can be reused to ensure the anonymity of arbitrarily many token transactions.

Generalising the operation $\Report':\Ms[_{2n}]\leadsto\OI^{t}\cup \set{\bot}$, for $k\ge2$ denote by $\Report^{(k)}:\Ms[_{k\tm n}]\leadsto\OI^{t}\cup \set{\bot}$ the following ($\sigma\in \Ms[_{k\tm n}]$ is the input state):
\itstart
  \item For $i=k, k-1\dc2$, sequentially perform $\Swap(\wtl{\sigma}|_{1\dc n,(i-1)n+1\dc in})$, where $\wtl{\sigma}|_{1\dc n,(i-1)n+1\dc in}$ denotes the qubits $1\dc n,(i-1)n+1\dc in$ of the input state after performing all the previous swap test; abort the operation and output ``$\bot$'' if any swap test returns ``$1$'' (report that cheating has been detected).
  \item For $\wtl{\sigma}|_{n+1\dc2n}$ denoting the qubits $n+1\dc2n$ of the input state after performing all the swap test, perform $\Report(\wtl{\sigma}|_{n+1\dc2n})$ and output the response.
\itend
Note that $\Report^{(2)}$ stands for the same operation as $\Report'$.
Extend the above definition for $\sigma\in \Ms[_{\ell\tm n}]$, $\ell>k$:\ in this case, let $\Report^{(k)}(\sigma)$ refer to applying the above operation to the first $k\tm n$ qubits of $\sigma$.

\clm[c_pat_chain]
{
  For $k\ge2$ and any $\sigma\in \Ms[_{k\tm n}]$,
  \m{
    \PR{\Report^{(k)}(\sigma) = \bot}
     \ge \PR{\Report^{(2)}(\sigma) = \bot}.
  }
}

In the above statement, the first $n$ qubits of the input state $\sigma$ represent the pattern token that is being used repeatedly towards anonymity verification, in a sequence of $k-1$ transactions.

\prfstart[\clmref{c_pat_chain}]

Let $\bk{\chi}\in \Ps[_{b+k\tm n}]$ for some $b\in \NN$ be a purification of $\sigma$, such that the latter equals the partial trace that preserves the last $k\tm n$ qubits of $\bk{\chi}[]$.

Then the probability of the event $[\Report^{(k)}(\sigma) = \bot]$ equals the probability of the outcome ``$\bot$'' in the following experiment:
\itstart
  \item Let $\bk{\wtl\chi} = \bk{\chi}$ -- call it the \e{input register}.
  \item For $i=k, k-1\dc2$:
  \itstart
    \item Perform the swap test on the qubits $b+1\dc b+n,b+(i-1)n+1\dc b+in$ of $\bk{\wtl\chi}[]$:
    \item if the answer is ``$1$'', then halt and return ``$\bot$'';
    \item otherwise, let $\bk{\wtl\chi}$ denote the new state of the input register.
  \itend
\itend

By \clmref{c_swap_chain}, the probability of the outcome ``$\bot$'' in the above experiment is at least the probability of that outcome in a similar experiment without the swap test that occurs before the last one:\ i.e., for $i=k, k-1\dc4, 2$.
The latter probability is, in turn, at least that in the even shorter experiment, where $i=k, k-1\dc5, 2$.
By induction, the probability is at least that in the experiment where the only performed swap test corresponds to $i=2$ -- which is equal to
\m{
  \PR{\Report^{(2)}(\sigma) = \bot},
}
as required.

\prfend

\sect[s_qua_sch]{A repetitive token scheme\\{\large Construction and analysis}}[Construction and analysis]

We are ready to implement a repetitive quantum token scheme with classical verification (\defiref{d_tok_rep}) and analyse its security and anonymity in various contexts.

\nconstr[constr_qua]{A quantum token scheme}
{
  Let $4|k$, $n\deq 2k$, $\NM\deq 2^{\dr k4}-1$, $m\deq k\tm2^k$, $\NT\deq 2^{\dr k2}$ and $t\deq 2k$.
  The operations $\Mint:\: \OI^{m}\to (\Ps[_{n}])^{\NM}$, $\Report:\: \Ps[_{n}]\leadsto \OI^{t}$ and $\Test:\: \OI^{m}\times \bigcup_{j=0}^{\NT-1}(\OI^t)^j\times \OI^{t} \to \TF$ are defined as
  \m{
    \Mint(S)&\deq \lf( \sum_{i\in [2^k]}\bk{i}{S\lf(\set{k\tm(i-1)+1\dc k\tm i}\rt)} \rt)^{\ox \NM};\\
    \Report(\bk{\phi})&\deq \txt{the outcome of measuring $\bk{\phi}$ in the computational basis};\\
    \Test(S, H, (I,R))&\deq \Cases{\tp}{if $S\lf(\set{k\tm(I-1)+1\dc k\tm I}\rt) = R$ and $(I,R) \nin H$;}{\bt}{otherwise.}
  }
}

\clm[c_qua_is_valid]
{
  \constrref{constr_qua} is a valid repetitive quantum token scheme that satisfies \defiref{d_tok_rep} with parameters $\combr \big(n=2k, m=k\tm2^{\dr k4}, \NM=2^{\dr k4}-1, \NT=2^{\dr k2}, t=k, \el=2^{-\dr k2}, \ef=6\tm 2^{-\dr k4}\big)$ for any $4|k$.
}

\prfstart

\constrref{constr_qua} satisfies the additional requirement of \defiref{d_tok_rep}, so it remains to see that it satisfies \defiref{d_tok} with parameters $\combr \big(n=2k, m=k\tm2^{\dr k4}, \NM=2^{\dr k4}-1, \NT=2^{\dr k2}, t=k, \el=2^{-\dr k2}, \ef=6\tm 2^{-\dr k4}\big)$ for any $4|k$.

To establish condition~\ref{d_tok_test_pass} of \defiref{d_tok}, we have to show that
\m{
  \PR[\Cl S, \Cl H, \Cl R]{\Test(\Cl S, \Cl H, \Cl R) = \tp} \ge 1-2^{-\dr k2}
}
with respect to $\Cl S = s\unin\OI^{m}$, $\Cl H$ being the outcome of certain measurement and $\Cl R = (r_1,r_2)$ being the outcome of measuring the token state
\m[m_is_valid_tok]{
  \sum_{i\in [2^k]}\bk{i}{s\lf(\set{k\tm(i-1)+1\dc k\tm i}\rt)}
}
in the computational basis.
We will prove a slightly stronger statement, namely that the above holds with respect to \e{any} value $\Cl H=h\in \OI^{k\tm j}$ ($j\in \set{0\dc 2^{\dr k2}-1}$) that may, in particular, depend on $s$ (though not on $(r_1,r_2)$).

By construction, $\Test(s, h, (r_1,r_2))$ accepts if and only if both $r_2 = s(\set{k\tm(r_1-1)+1\dc k\tm r_1})$ and $(r_1,r_2) \nin h$.
The former holds with certainty; the latter fails only if $r_1\unin [2^k]$ is among the indices that are contained in $h$ -- at most, $j<2^{\dr k2}$ of them.
Accordingly,
\m{
  \PR{\Test(s, h, (r_1,r_2)) = \tp} > 1-2^{-\dr k2} = 1-\el,
}
as required.

Towards condition~\ref{d_tok_fraud} of \defiref{d_tok}, observe that every instance of the token state \bref{m_is_valid_tok} can be generated via a single quantum query with respect to $F_s:\: i\to s\lf(\set{k\tm(i-1)+1\dc k\tm i}\rt)$, in which case the choice $s\unin\OI^{m}$ corresponds to a uniformly random $F_s:\: [2^k]\to[2^k]$.
Then it follows from \lemref{l_ququ} that a list of at most $N$ distinct pairs $(x_i,y_i)$ such that $\sz{\set{i}[y_i=F_s(x_i)]} \ge q+1$ can be produced by making at most $q$ queries to $F_s$ -- that is, via measuring at most $q$ instances of the token state~\bref{m_is_valid_tok} -- with probability at most
\m[m_is_valid_fraud]{
  6N\tm \fr{q+1}{2^k}.
}

In terms of \defiref{d_tok}, the measurement \Fraud\ measures $q=\sz{s_0}\le \NM=2^{\dr k4}-1$ instances of the token state~\bref{m_is_valid_tok} and outputs $\NT=2^{\dr k2}$ pairs.
For $[\sz[]{\bTest(\Cl S, \Cl H)}>\sz{s_0}]$ to hold, it must be the case that at least $\sz{s_0}+1$ of those $2^{\dr k2}$ pairs are distinct ones of the form $(i,F_s(i))$ -- according to~\bref{m_is_valid_fraud}, this happens with probability at most $\ef=6\tm 2^{-\dr k4}$, as required.

\prfend[\clmref{c_qua_is_valid}]



% ...



% ...

\sect[s_app]{Applications}

Below we discuss a number of direct applications of the tokens defined in Construction \ref{constr_qua}.

\ssect[ss_an_money]{Anonymous quantum money}

We can use our construction for one-time payments, which evade counterfeiting, use classical communication, and are anonymous. The first two properties were achieved by Ben-David and Sattath~\cite[Section 8.1]{Ben-David2023-in}, but their construction did not achieve anonymity. This use case proceeds as follows:
\begin{enumerate}
	\item The bank mints tokens using $\Mint()$, and sends subsets of the tokens to users using quantum communication. 
    \item Users can swap tokens with each other to check if they are being tracked by the bank, as in Claim \ref{c_rep_ind}.
	\item To redeem a token, a user first measures it by applying $\Report()$ and stores the result in classical memory. This can be done immediately upon reception of the token to avoid decoherence or expelling quantum information processing resources to dynamically correct the quantum memory.
	\item At any time of their choosing, the user can send the result of their measurement $(I,R)$ to the bank using classical communication.
	\item  The bank keeps a history $H$ of redeemed tokens, and accepts the token if $\Test(S,H,(I,R))$ passes. The bank then adds $(I,R)$ to $H$.
\end{enumerate}

The requirement of using quantum communication only during token distribution can be thought of as analogous to drawing cash from an ATM. One must use a device controlled by the bank, in which case it may be more acceptable to use quantum communication. Beyond this point, only classical memory is required. 

Correctness and security follow directly from that of the token scheme in Construction \ref{constr_qua}.

\ssect[ss_an_pub]{Anonymous one-time pads}

The strings generated by measuring the tokens (using $\Report()$) are random and with high probability unique to each token, which implies that they can be used directly in the one-time pad (OTP) protocol, as outlined below:

\begin{enumerate}
	\item The bank mints tokens using $\Mint()$, and sends subsets of the tokens to users using quantum communication.
    \item Users can audit the bank as in Claim \ref{c_rep_ind}.
	\item Users measure each token using $\Report()$ and store the result in classical memory.
	\item Denoting by $(I,R)$ the result of $\Report(\ket{\phi})$ for a token $\ket{\phi}$, the user then encodes a message $M$ of $k$ bits by computing $C=R \oplus M$ and sends $(I,C)$ to the bank.
    \item The bank decodes the encoded message by computing $C\oplus S\lf(\set{k\tm(I-1)+1\dc k\tm I}\rt)$.
\end{enumerate}

Multiple tokens can be used to encode longer messages, and the security and correctness follow immediately from those of the original token scheme. Security here implies that one user (or an outsider) cannot decode a message encoded by another user, since this would imply that the corresponding token can be counterfeited. Since the bank issued identical tokens to all users, it has no knowledge of the identity of the sender of the encrypted message, despite having the ability to decrypt it.  

\ssect[ss_an_vot]{Anonymous voting}

The application of anonymous quantum money schemes to voting was first shown by~\cite{CGY24}. Voting can be reduced to sending messages using one-time pads, where each voter receives a single token and uses it to encrypt their vote. Uncloneability of tokens prevents double voting, and the anonymity guarantees ensure that the vote administrator cannot associate users with their votes without this being detected. 

% ...


Previous proposals for quantum voting systems~\cite{Vaccaro2007-jn, Sun2019-uk} require that the vote itself be a quantum message between the voter and the bank, and some rely on a quantum blockchain, which is a more involved setup in various ways~\cite{Sun2019-uk}. There are also classical voting schemes based on distributing OTPs between all pairs of players~\cite{Chaum1988-fx}. This scheme provides weaker privacy guarantees that do not detect misbehavior of the vote administrator (i.e., the bank).  

\sect[s_discussion]{Discussion}

We have presented a quantum token scheme that exhibits unconditional security, enables classical verification,  and enables anonymity by empowering users to check whether they are being tracked by the issuing authority. One drawback of our approach is that only the bank can check the validity of a token, in contrast to so-called public key quantum money schemes, where any user can efficiently verify quantum money. While the initial proposals for public key quantum money schemes required oracles~\cite{Aaronson2009-uz, Aaronson2012-bg} or proved difficult to construct securely~\cite{Zhandry2019-ru, Roberts2021-vx}, a recent breakthrough~\cite{Shmueli2025-vi} shows how to construct even more powerful objects known as one-shot signatures~\cite{Amos2020-ls} from standard assumptions. It may thus be interesting to consider whether public key analogs of our construction can be obtained based on these results. One-shot signatures can even be used to construct public key quantum money with classical communication alone~\cite{Amos2020-ls}, improving upon our need for a single round of quantum communication, yet the signature chains required to implement such a scheme might be hard to reconcile with our notion of anonymity. 

We call our construction quantum tokens due to their single-use nature (following~\cite{Bennett1983-jx, Ben-David2023-in}). We restrict ourselves to this setting since it may be challenging to analyze the security of the scheme if the tokens are allowed to change hands multiple times. Extending our results to account for this is an interesting direction for future work.

We make no attempt to address computational efficiency in this work. The efficiency of our construction could be improved by replacing the perfect randomness we assume with pseudorandomness, albeit at the cost of introducing computational assumptions.


\sect*{Acknowledgments}

The authors are grateful to Ronald de Wolf and Scott Aaronson for insightful discussions and most useful references. DG was \thanksDG~SJ was partially funded by Scott Aaronson's CIQC grant.

\newpage 
\toct{References}

\bibliographystyle{alpha}

\bibliography{tex, paperpile}

\end{document}

%% file: dgB.tex
\usepackage{amssymb}
\usepackage{amsmath}%

\usepackage{amsthm}
\theoremstyle{definition} 

\usepackage{hyperref}

\usepackage{units}%

\usepackage{graphicx}

\usepackage{xcolor}

\usepackage{marginnote}%

\usepackage{mathrsfs}%

\usepackage{soul}%
\setul{1.25pt}{.16pt}%

\usepackage{newunicodechar}%
\newunicodechar{⇒}{\ensuremath{\Rightarrow}}
\newunicodechar{ꞏ}{\scalebox{0.75}{\ensuremath{\bullet}}}

\usepackage[T1,OT2,T2A]{fontenc}

\usepackage[utf8]{inputenc}

\usepackage[shorthands=off,greek,czech,ukrainian,english]{babel}

\let\sq\relax%

\pdfpageheight=\paperheight
\pdfpagewidth=\paperwidth

\makeatletter

\newcommand{\NewcommandThreeArgsTwoOpt}[6][*]
{
\ifthenelse{\equal{}{#1}}
{\def\comm@TMP{\DeclareRobustCommand}}
{\def\comm@TMP{\newcommand}}
\comm@TMP{#2}{\@ifnextchar[%
{\csname\expandafter\@gobble\string#2@presq\endcsname}%
{\csname\expandafter\@gobble\string#2@nopresq\endcsname}}
\expandafter\def\csname\expandafter\@gobble\string#2@nopresq\endcsname##1{\@ifnextchar[%
{\csname\expandafter\@gobble\string#2@nopresq@postsq\endcsname[]{##1}}%
{\csname\expandafter\@gobble\string#2@nopresq@nopostsq\endcsname[]{##1}}}
\expandafter\def\csname\expandafter\@gobble\string#2@presq\endcsname[##1]##2{\@ifnextchar[%
{\csname\expandafter\@gobble\string#2@presq@postsq\endcsname[{##1}]{##2}}%
{\csname\expandafter\@gobble\string#2@presq@nopostsq\endcsname[{##1}]{##2}}}
\expandafter\def\csname\expandafter\@gobble\string#2@nopresq@nopostsq\endcsname[##1]##2{#3}
\expandafter\def\csname\expandafter\@gobble\string#2@presq@nopostsq\endcsname[##1]##2{#4}
\expandafter\def\csname\expandafter\@gobble\string#2@nopresq@postsq\endcsname[##1]##2[##3]{#5}
\expandafter\def\csname\expandafter\@gobble\string#2@presq@postsq\endcsname[##1]##2[##3]{#6}
}

\newcommand{\NewcommandTwoArgsLastOpt}[4][*]
{
\ifthenelse{\equal{}{#1}}
{\def\comm@TMP{\DeclareRobustCommand}}
{\def\comm@TMP{\newcommand}}
\comm@TMP{#2}{\csname\expandafter\@gobble\string#2@def\endcsname}
\expandafter\def\csname\expandafter\@gobble\string#2@def\endcsname##1{\@ifnextchar[%
{\csname\expandafter\@gobble\string#2@postsq\endcsname{##1}}%
{\csname\expandafter\@gobble\string#2@nopostsq\endcsname{##1}}}
\expandafter\def\csname\expandafter\@gobble\string#2@nopostsq\endcsname##1{#3}
\expandafter\def\csname\expandafter\@gobble\string#2@postsq\endcsname##1[##2]{#4}
}

\newcommand{\NewcommandFourArgsTwoOpt}[6][*]
{
\ifthenelse{\equal{}{#1}}
{\def\comm@TMP{\DeclareRobustCommand}}
{\def\comm@TMP{\newcommand}}
\comm@TMP{#2}{\@ifnextchar[%
{\csname\expandafter\@gobble\string#2@presq\endcsname}%
{\csname\expandafter\@gobble\string#2@nopresq\endcsname}}
\expandafter\def\csname\expandafter\@gobble\string#2@nopresq\endcsname##1##2{\@ifnextchar[%
{\csname\expandafter\@gobble\string#2@nopresq@postsq\endcsname[]{##1}{##2}}%
{\csname\expandafter\@gobble\string#2@nopresq@nopostsq\endcsname[]{##1}{##2}}}
\expandafter\def\csname\expandafter\@gobble\string#2@presq\endcsname[##1]##2##3{\@ifnextchar[%
{\csname\expandafter\@gobble\string#2@presq@postsq\endcsname[{##1}]{##2}{##3}}%
{\csname\expandafter\@gobble\string#2@presq@nopostsq\endcsname[{##1}]{##2}{##3}}}
\expandafter\def\csname\expandafter\@gobble\string#2@nopresq@nopostsq\endcsname[##1]##2##3{#3}
\expandafter\def\csname\expandafter\@gobble\string#2@presq@nopostsq\endcsname[##1]##2##3{#4}
\expandafter\def\csname\expandafter\@gobble\string#2@nopresq@postsq\endcsname[##1]##2##3[##4]{#5}
\expandafter\def\csname\expandafter\@gobble\string#2@presq@postsq\endcsname[##1]##2##3[##4]{#6}
}

\newcommand{\NewcommandFourArgsTwoOptDefa}[5][*]
{
\ifthenelse{\equal{}{#1}}
{\def\comm@TMP{\NewcommandFourArgsTwoOpt[]}}
{\def\comm@TMP{\NewcommandFourArgsTwoOpt}}
\expandafter\newcommand\csname\expandafter\@gobble\string#2@full\endcsname[4]{#5}
\comm@TMP{#2}
{\csname\expandafter\@gobble\string#2@full\endcsname{#3}{##2}{##3}{#4}}
{\csname\expandafter\@gobble\string#2@full\endcsname{##1}{##2}{##3}{#4}}
{\csname\expandafter\@gobble\string#2@full\endcsname{#3}{##2}{##3}{##4}}
{\csname\expandafter\@gobble\string#2@full\endcsname{##1}{##2}{##3}{##4}}
}

\newcommand\bk{\@ifnextchar[%
{\lf\langle \bk@bra{}}%
{\@ifnextchar<%
{\lf.\bk@op}
{\lf| \bk@ket{}}%
}%
}

\def\bk@bra#1[#2]{\def\bk@tmp{#2}\ifx\bk@tmp\empty{#1}\else{#2}\fi\@ifnextchar[%
{\md|\hspace{-1.5pt}\md\langle \bk@bra{\def\bk@tmp{#2}\ifx\bk@tmp\empty{#1}\else{#2}\fi}}%
{\@ifnextchar\bgroup%
{\md| \bk@ket{\def\bk@tmp{#2}\ifx\bk@tmp\empty{#1}\else{#2}\fi}}%
{\@ifnextchar<%
{\md| \bk@op}%
{\rt|}%
}%
}%
}

\def\bk@ket#1#2{\def\bk@tmp{#2}\ifx\bk@tmp\empty{#1}\else{#2}\fi\@ifnextchar[%
{\md\rangle\hspace{-1.5pt}\md\langle \bk@bra{\def\bk@tmp{#2}\ifx\bk@tmp\empty{#1}\else{#2}\fi}}%
{\@ifnextchar\bgroup%
{\md\rangle\hspace{-1.5pt}\md| \bk@ket{\def\bk@tmp{#2}\ifx\bk@tmp\empty{#1}\else{#2}\fi}}%
{\@ifnextchar<%
{\md\rangle \bk@op}%
{\rt\rangle}%
}%
}%
}

\def\bk@op<#1>{#1 \@ifnextchar[%
{\md\langle \bk@bra{}}%
{\@ifnextchar\bgroup%
{\md| \bk@ket{}}%
{\@ifnextchar<%
{\tm \bk@op}%
{\rt.}%
}%
}%
}

\newcommand\Cases[1][0pt]{%
\def\Case@skipam{#1}%
\left\{\!\!\!\begin{array}{ll}\Cases@continue%
}

\def\Cases@continue#1#2{\@ifnextchar\bgroup%
{#1 &\txt{#2}\\[\Case@skipam] \Cases@continue}%
{#1 &\txt{#2}\end{array}\Cases@end}%
}

\def\Cases@end{\@ifnextchar[%
{\Cases@end@postsq}%
{\right.}
}

\def\Cases@end@postsq[#1]{\ifthenelse{\equal{\}}{#1}}
{\!\!\right\}}%
{\right.}
}

\newcommand{\MyMakeTheoMacros}[3]{
\expandafter\newcommand\csname\expandafter\@gobble\string#2NostarNoname@DGaux\endcsname[2][]
{\ifthenelse{\equal{}{##1}}%
{\begin{#1}~##2 \end{#1}}%
{\begin{#1}\label{##1}~##2\end{#1}}%
}
\expandafter\newcommand\csname\expandafter\@gobble\string#2StarNoname@DGaux\endcsname[1]
{\begin{#1*}~##1 \end{#1*}}
\newcommand#2{\expandafter\@ifstar%
\expandafter{\csname\expandafter\@gobble\string#2StarNoname@DGaux\endcsname}%
{\csname\expandafter\@gobble\string#2NostarNoname@DGaux\endcsname}%
}

\expandafter\newcommand\csname\expandafter\@gobble\string#2NostarName@DGaux\endcsname[3][]
{\ifthenelse{\equal{}{##1}}%
{\begin{#1}[\e{##2}]~##3 \end{#1}}%
{\begin{#1}[\e{##2}]\label{##1}~##3\end{#1}}%
}
\expandafter\newcommand\csname\expandafter\@gobble\string#2StarName@DGaux\endcsname[2]
{\begin{#1*}[\e{##1}]~##2 \end{#1*}}
\newcommand#3{\expandafter\@ifstar%
\expandafter{\csname\expandafter\@gobble\string#2StarName@DGaux\endcsname}
{\csname\expandafter\@gobble\string#2NostarName@DGaux\endcsname}%
}
}

\DeclareRobustCommand\dgref[2]{\@ifnextchar[%
{#2\dgref@presq}%
{#1\dgref@nopresq}%
}

\def\dgref@presq[#1]{\@ifnextchar[%
{\ref{#1}, \dgref@presq}
{\ref{#1} and~\dgref@nopresq}
}

\def\dgref@nopresq#1{\ref{#1}}

\DeclareRobustCommand\bref{\@ifnextchar[%
{\bref@presq}%
{\bref@nopresq}%
}

\def\bref@presq[#1]{\@ifnextchar[%
{(\ref{#1}), \bref@presq}
{(\ref{#1}) and~\bref@nopresq}
}

\def\bref@nopresq#1{(\ref{#1})}

\newcommand{\ifndef}[2]{\@ifundefined{#1}{#2}{}}

\makeatother

\newcommand{\NewELike}[2]{
\NewcommandThreeArgsTwoOpt{#1}
{#2\lf[##2\rt]}
{#2_{##1}\lf[##2\rt]}
{#2\lf[##2\md|{\ud}##3\rt]}
{#2_{##1}\lf[##2\md|{\ud}##3\rt]}
}

\NewELike{\PR}{\mathop{\pmb{Pr}}}

\newcommand{\NewMinLike}[2]{
\NewcommandThreeArgsTwoOpt{#1}
{#2\lf\{##2\rt\}}
{#2_{##1}\lf\{##2\rt\}}
{#2\lf\{##2\md|{\ud}##3\rt\}}
{#2_{##1}\lf\{##2\md|{\ud}##3\rt\}}
}

\NewMinLike{\Min}{min}

\newcommand{\NewOp}[2]{
\newcommand{#1}[2][]{\ensuremath{\operatorname{#2}##1\ifthenelse{\equal{}{##2}}{}{\lf(##2\rt)}}}
}

\newcommand{\NewOpI}[2]{\NewOp{#1}{\mathit{#2}}}

\NewOp{\tr}{tr}

\NewOpI{\aso}{o}

\newcommand{\NewId}[2]{
\newcommand{#1}[1][]{\ensuremath{\ifthenelse{\equal{}{##1}}{\operatorname{\mathit{#2}}}{\operatorname{\mathit{#2##1}}}}}
}

\newcommand{\NewIdR}[2]{
\newcommand{#1}[1][]{\ensuremath{\ifthenelse{\equal{}{##1}}{\operatorname{\mathrm{#2}}}{\operatorname{\mathrm{#2##1}}}}}
}

\newcommand{\NewIdSm}[2]{\NewId{#1}%
{\mathchoice{#2}{#2}{\scriptscriptstyle{#2}}{\scalebox{0.75}{$\scriptscriptstyle{#2}$}}}%
}

\NewIdSm{\tp}{\top}
\NewIdSm{\bt}{\bot}

\NewIdSm{\orth}{\perp}

\newcommand{\OI}{\set{0,1}}
\newcommand{\TF}{\set{\tp,\bt}}

\newcommand{\Then}{\Longrightarrow}

\newcommand{\MyMakeRefMacros}[3]{\newcommand{#1}{\dgref{#2}{#3}}}

\ifndef{theorem}{}

\MyMakeTheoMacros{theorem}{\theo}{\ntheo}

\MyMakeRefMacros{\theoref}{Theorem~}{Theorems~}

\ifndef{definition}{}

\MyMakeTheoMacros{definition}{\defi}{\ndefi}

\MyMakeRefMacros{\defiref}{Definition~}{Definitions~}

\ifndef{claim}{}

\MyMakeTheoMacros{claim}{\clm}{\nclm}

\MyMakeRefMacros{\clmref}{Claim~}{Claims~}

\ifndef{lemma}{}

\MyMakeTheoMacros{lemma}{\lem}{\nlem}

\MyMakeRefMacros{\lemref}{Lemma~}{Lemmas~}

\ifndef{corollary}{}

\MyMakeTheoMacros{corollary}{\crl}{\ncrl}

\MyMakeRefMacros{\crlref}{Corollary~}{Corollaries~}

\ifndef{fact}{}

\MyMakeTheoMacros{fact}{\fct}{\nfct}

\MyMakeRefMacros{\fctref}{Fact~}{Facts~}

\ifndef{question}{}

\MyMakeTheoMacros{question}{\quest}{\nquest}

\MyMakeRefMacros{\questref}{Question~}{Questions~}

\ifndef{construction}{}

\MyMakeTheoMacros{construction}{\constr}{\nconstr}

\MyMakeRefMacros{\constrref}{Construction~}{Constructions~}

\renewcommand{\qedsymbol}{$\blacksquare$}

\newcommand{\prfstart}[1][]{\ifthenelse{\equal{}{#1}}%
{\begin{proof}[\underline{\proofname}]\renewcommand{\qedsymbol}{$\blacksquare$}}%
{\begin{proof}[\underline{\proofname\ #1}]%
\renewcommand{\qedsymbol}{$\blacksquare_{\mbox{\it{\scriptsize{#1}}}}$}}%
}

\newcommand{\prfend}[1][*]{%
\ifthenelse{\equal{}{#1}}{\renewcommand{\qedsymbol}{$\blacksquare$}}{}%
\ifthenelse{\equal{*}{#1}}{}%
{\renewcommand{\qedsymbol}{$\blacksquare_{\mbox{\it{\scriptsize{#1}}}}$}}%
\end{proof}\renewcommand{\qedsymbol}{$\blacksquare$}%
}

\newcommand{\abstart}{\begin{abstract}}
\newcommand{\abend}{\end{abstract}}

\newcommand{\itstart}[1][]{%
\ifthenelse{\equal{}{#1}}%
{\begin{itemize}}%
{\begin{itemize}[#1]}%
}
\newcommand{\itend}{\end{itemize}}

\newcommand{\enstart}[1][]{%
\ifthenelse{\equal{}{#1}}%
{\begin{enumerate}}%
{\begin{enumerate}[#1]}%
}
\newcommand{\enend}{\end{enumerate}}

\newcommand{\NewSectLike}[3]{
\NewcommandThreeArgsTwoOpt{#1}
{\ifthenelse{\equal{*}{##2}}{#2*}{#2{##2}}}
{#2{##2\label{##1}}}
{#2[##3]{##2#3{##3}}#3{##3}}
{#2[##3]{##2#3{##3}\label{##1}}#3{##3}}
}

\NewSectLike{\sect}{\section}{\sectionmark}

\NewSectLike{\ssect}{\subsection}{\subsectionmark}

\NewSectLike{\sssect}{\subsubsection}{\subsubsectionmark}

\MyMakeRefMacros{\sref}{Section~}{Sections~}
\MyMakeRefMacros{\ssref}{Subsection~}{Subsections~}

\newcommand{\toc}[3][]{%
\ifthenelse{\equal{chapter}{#2}}{\newpage}{}%
\ifthenelse{\equal{part}{#2}}{\newpage}{}%
\ifthenelse{\equal{}{#1}}%
{\phantomsection\addcontentsline{toc}{#2}{#3}}%
{\phantomsection\refstepcounter{#2}\label{#1}\addcontentsline{toc}{#2}{#3}}%
}

\newcommand{\toct}[1]{\toc{section}{#1}}

\newcommand{\fn}[2][]{%
\IfMathMode{\MyComment{Footnotes may not work here!}}{}%
\ifthenelse{\equal{}{#1}}%
{\footnote{
\ignorespaces #2\unskip}}%
{\footnote{\label{#1}
\ignorespaces #2}}%
}

\newcommand{\fnm}{\footnotemark}
\newcommand{\fnt}[2][]{\ifthenelse{\equal{}{#1}}%
{\footnotetext{
\ignorespaces #2}\noindent\ignorespaces}%
{\footnotetext{\label{#1}
\ignorespaces #2}\noindent\ignorespaces}%
}

\MyMakeRefMacros{\fnref}{Footnote~}{Footnotes~}

\newcommand{\MyComment}[1]{\ClassWarning{My Macros}{#1}}

\newcommand{\IfMathMode}[2]{\ifmmode{#1}\else{#2}\fi}

\newcommand{\m}[2][]{%
\ifthenelse{\boolean{delim_grow}}{\delimiterfactor=1001}{}%
\ifthenelse{\equal{}{#1}}%
{\begin{align*} #2 \end{align*}}%
{\ifthenelse{\equal{P}{#1}}%
{\allowdisplaybreaks\begin{align*} #2%
\end{align*}\interdisplaylinepenalty=10000}%
{\ifthenelse{\equal{I}{#1}}%
{$#2$}%
{\begin{align}\begin{split}\label{#1} #2 \end{split}\end{align}}%
}%
}%
}

\makeatletter

\let\dgampersand\&

\DeclareRobustCommand\&{%
\new@ifnextchar[%
{\dgsep@reposit}%
{\dgampersand}%
}

\def\dgsep@reposit[#1]{\hspace{#1}&\hspace{-#1}}

\makeatother

\mathchardef\breakingcomma\mathcode`\,
{\catcode`,=\active
\gdef,{\breakingcomma\discretionary{}{}{}}
}

\newcommand{\combr}{\mathcode`\,=\string"8000}

\newcommand{\lf}{\mathopen{}\mathclose\bgroup\left}
\newcommand{\rt}{\aftergroup\egroup\right}

\providecommand{\middle}{\big}
\newcommand{\md}{\middle}

\newcommand{\ud}{\vphantom{|_1^1}}

\newcommand{\NewPar}[4][*]{
\ifthenelse{\equal{}{#1}}
{\def\comm@TMP{\NewcommandThreeArgsTwoOpt[]}}
{\def\comm@TMP{\NewcommandThreeArgsTwoOpt}}
\comm@TMP{#2}
{\lf#3{##2}\rt#4}
{#3{##2}#4}
{\lf#3{##2}\rt#4_{\!##3}}
{#3{##2}#4_{\!##3}}
}

\NewPar{\lrp}{(}{)}
\NewPar{\lra}{\langle}{\rangle}
\NewPar[]{\sz}{|}{|}
\NewPar[]{\norm}{\|}{\|}
\NewPar[]{\ceil}{\lceil}{\rceil}
\NewPar[]{\floor}{\lfloor}{\rfloor}
\NewPar{\round}{\lfloor}{\rceil}

\NewcommandThreeArgsTwoOpt[]{\set}
{\lf\{#2\rt\}}
{\MyComment{Second optional argument missing in "\set"?}}
{\lf\{#2\md|\ud#3\rt\}}
{\lf\{#2\ud#1\ud#3\rt\}}

\newcommand{\chs}[2]{\lf(\!\genfrac{}{}{0cm}{}{#1}{#2}\!\rt)}%

\definecolor{Red}{rgb}{1,0,0}
\definecolor{DarkRed}{rgb}{0.54,0,0}

\DeclareTextFontCommand{\bemph}{\bfseries}
\DeclareTextFontCommand{\ibemph}{\bfseries\em}
\DeclareTextFontCommand{\ttl}{\ttfamily}

\newcommand{\e}{\emph}

\newcommand{\Cl}{\mathcal}%

\newcommand{\txt}[1]{\textrm{#1}}%

\newcommand{\eps}{\varepsilon}

\newcommand{\fr}{\frac}

\providecommand{\NN}{\mathbb{N}}
\providecommand{\CC}{\mathbb{C}}
\providecommand{\RR}{\mathbb{R}}

\newcommand{\tm}{\cdot}
\newcommand{\ox}{\otimes}%
\newcommand{\sq}{\sqrt}

\newcommand{\dr}{\nicefrac}

\newcommand{\ds}[1][]
{\ifthenelse{\equal{}{#1}}{\allowbreak\ldots}{#1\allowbreak\ldots#1}}
\newcommand{\dc}{\ds[,]}

\newcommand{\teq}{\triangleq}
\newcommand{\deq}{\stackrel{\textrm{def}}{=}}

\newcommand{\nin}{\not\in}%

\NewcommandFourArgsTwoOptDefa{\overlay}
{0mu}{1}
{
\begingroup
\mathchoice{%
\bgroup
\scalebox{#4}{\ooalign{$\displaystyle#2\mkern#1$\cr
\hidewidth{$\displaystyle\mkern#1#3$}\hidewidth}}
\egroup
}{%
\bgroup
\scalebox{#4}{\ooalign{$\textstyle#2\mkern#1$\cr
\hidewidth{$\textstyle\mkern#1#3$}\hidewidth}}
\egroup
}{%
\bgroup
\scalebox{#4}{\ooalign{$\scriptstyle#2\mkern#1$\cr
\hidewidth{$\scriptstyle\mkern#1#3$}\hidewidth}}
\egroup
}{%
\bgroup
\scalebox{#4}{\ooalign{$\scriptscriptstyle#2\mkern#1$\cr
\hidewidth{$\scriptscriptstyle\mkern#1#3$}\hidewidth}}
\egroup
}
\endgroup
}

\newcommand{\unin}{\mathrel{\overlay[1.25mu]{\subset}{\sim}}}

\newcommand{\sbseq}{\subseteq}
\newcommand{\sbs}{\subset}

\NewcommandTwoArgsLastOpt{\GF}
{{\mathcal{G\!F}_{#1}}}
{{\mathcal{G\!F}_{#1}^{#2}}}

\newcommand*{\wht}{\widehat}
\newcommand*{\wtl}{\widetilde}

\newcommand{\wbr}[1]{\mskip.5\thinmuskip\overline{\mskip-.5\thinmuskip {#1} \mskip-.5\thinmuskip}\mskip.5\thinmuskip}%

\def\?{\mskip 0.5\thinmuskip}%

%% file: tikz_figure.tex
\definecolor{purple}{RGB}{128,0,128}
\definecolor{grey}{RGB}{128,128,128}
\definecolor{orange}{RGB}{255,165,0}

\tikzset{
    box/.style={
        rectangle, rounded corners, thick, draw=#1, text centered,
        minimum height=2cm, minimum width=4cm, fill=#1!20,
        text width=3.5cm, align=center, font=\bfseries, drop shadow,
    },
    qubit/.style={
        circle, draw=black, fill=white, very thick, minimum size=0.5cm, font=\bfseries,
    },
    classical/.style={
        rectangle, rounded corners, draw=grey, fill=white, very thick,
        minimum size=0.5cm, font=\bfseries,
    },
    midarrow/.style={
        decoration={markings, mark=at position 0.55 with {\arrow[scale=1.5]{latex}}},
        postaction={decorate}
    },
    midarrowbig/.style={
        decoration={markings, mark=at position 0.6 with {\arrow[scale=2.5]{latex}}},
        postaction={decorate}
    },
    midarrowbig1/.style={
        decoration={markings, mark=at position 0.7 with {\arrow[scale=2.5]{latex}}},
        postaction={decorate}
    },
    midarrowbig2/.style={
        decoration={markings, mark=at position 0.75 with {\arrow[scale=2.5]{latex}}},
        postaction={decorate}
    },
    quantum arrow/.style={
        decorate, decoration={snake, amplitude=0.5pt, segment length=6pt},
    },
    note/.style={
        font=\small, text width=3cm, align=center, fill=white,
        draw=gray, rounded corners, inner sep=4pt, drop shadow,
    },
    actor/.style={
        text centered, text width=4cm, 
        font=\bfseries,
    },
}

\begin{tikzpicture}

\coordinate (bank_col) at (0,0);
\coordinate (userq_col) at (4.5,0); 
\coordinate (userc_col) at (9,0);   
\coordinate (row_top) at (0,6.0);      
\coordinate (row_mint) at (0,5.1);    
\coordinate (row_audit) at (0,3.4);     
\coordinate (row_storage) at (0,1.7); 
\coordinate (row_redeem) at (0,0);    
\coordinate (row_labels) at (-3,0); 

\node[actor] (bank_label) at (bank_col |- row_top) {\small Bank \\ Classical Memory};
\node[actor] (userq_label) at (userq_col |- row_top) {\small Users \\ Quantum Memory};
\node[actor] (userc_label) at (userc_col |- row_top) {\small Users \\ Classical Memory};
\node[font=\bfseries] (minting) at (row_labels |- row_mint) {Minting};
\node[font=\bfseries] (auditing) at (row_labels |- row_audit) {Auditing};
\node[font=\bfseries, below=0.1cm of auditing] {(optional)};
\node[font=\bfseries] (storage) at (row_labels |- row_storage) {Redemption};
\node[font=\bfseries] (redemption) at (row_labels |- row_redeem) {Verification};

\node[classical] (bank_mint) at (bank_col |- row_mint) {$(S,H)$};
\node[qubit, fill=purple, fill opacity=0.5, text opacity=1] (userq_mint1) at ($(userq_col |- row_mint) - (0.75,0)$) {$\ket{\phi}$};
\node[qubit, fill=purple, fill opacity=0.5, text opacity=1] (userq_mint2) at ($(userq_col |- row_mint) + (0.75,0)$) {$\ket{\phi}$};
\draw[quantum arrow, purple, very thick] (bank_mint) to (userq_mint1);
\draw[draw=none, midarrowbig, purple] (bank_mint) to (userq_mint1); 
\draw[quantum arrow, purple, very thick] (bank_mint) to[bend right=30] (userq_mint2);
\draw[draw=none, midarrowbig, purple] (bank_mint) to[bend right=30] (userq_mint2); 

\node[classical] (bank_audit) at (bank_col |- row_audit) {$(S,H)$};
\node[qubit, fill=purple, fill opacity=0.5, text opacity=1] (userq_audit1) at ($(userq_col |- row_audit) - (0.75,0)$) {$\ket{\phi}$};
\node[qubit, fill=purple, fill opacity=0.5, text opacity=1] (userq_audit2) at ($(userq_col |- row_audit) + (0.75,0)$) {$\ket{\phi}$};
\node[] (userq_audit2_a) at ($(userq_audit2) + (0,0.55)$) {};
\node[] (userq_audit1_a) at ($(userq_audit1) + (0,0.55)$) {};
\node[] (userq_audit2_b) at ($(userq_audit2) - (0,0.55)$) {};
\node[] (userq_audit1_b) at ($(userq_audit1) - (0,0.55)$) {};
\draw[quantum arrow, purple, very thick] (userq_audit1) to [bend left=45] (userq_audit2);
\draw[draw=none, midarrowbig1, purple] (userq_audit1_a) to (userq_audit2_a); 
\draw[draw=none, midarrowbig2, purple] (userq_audit2_b) to (userq_audit1_b); 
\draw[quantum arrow, purple, very thick] (userq_audit2) to [bend left=45] (userq_audit1);

\node[classical] (bank_storage) at (bank_col |- row_storage) {$(S,H)$};
\node[qubit, fill=purple, fill opacity=0.5, text opacity=1] (userq_storage) at ($(userq_col |- row_storage) - (0.75,0)$) {$\ket{\phi}$};
\node[qubit, fill=purple, fill opacity=0.5, text opacity=1] (userq_storage2) at ($(userq_col |- row_storage) + (0.75,0)$) {$\ket{\phi}$};
\node[classical, draw=grey] (userc_storage1) at ($(userc_col |- row_storage) - (1,0)$) {$(I_1,R_1)$};
\node[classical, draw=grey] (userc_storage2) at ($(userc_col |- row_storage) + (1,0)$) {$(I_2,R_2)$};
\draw[orange, very thick, midarrow] (userq_storage) to[bend right=-30] (userc_storage1);
\draw[orange, very thick, midarrow] (userq_storage2) to[bend right=30] (userc_storage2);

\node[classical, draw=grey] (bank_redemption) at (bank_col |- row_redeem) {$(S,H)$};
\node[classical, draw=grey] (userc_redeem1) at ($(userc_col |- row_redeem) - (1,0)$) {$(I_1,R_1)$};
\node[classical, draw=grey] (userc_redeem2) at ($(userc_col |- row_redeem) + (1,0)$) {$(I_2,R_2)$};
\draw[grey, very thick, midarrow] (userc_redeem1) to (bank_redemption);
\draw[grey, very thick, midarrow] (userc_redeem2) to [bend right=15] (bank_redemption);

\end{tikzpicture}